\documentstyle[10pt,fullpage,psfig,bbbold,twocolum]{article}

\def\slantfrac#1#2{\hbox{$\,^{#1}\!/_{#2}$}}

\begin{document}

\title{Structure of gaseous flows in semidetached binaries after
mass transfer termination}

\author{D.V.Bisikalo$^1$, A.A.Boyarchuk$^1$,
A.A.Kilpio$^1$,\\
O.A.Kuznetsov$^2$, V.M.Chechetkin$^2$\\[0.3cm]
$^1$ {\it Institute of Astronomy of the Russian Acad. of Sci.,
Moscow}\\
{\sf bisikalo@inasan.rssi.ru; aboyar@inasan.rssi.ru;
skilpio@inasan.rssi.ru}\\[0.3cm]
$^2$ {\it Keldysh Institute of Applied Mathematics, Moscow}\\
{\sf kuznecov@spp.keldysh.ru; chech@int.keldysh.ru}\\[0.3cm]}
\date{}
\maketitle

\begin{abstract}

The results of three-dimensional numerical simulations of mass
transfer in semidetached binaries after mass transfer
termination is presented and the structure of residual accretion
disk was investigated. It was shown that after the mass transfer
termination the quasi-elliptic accretion disk becomes circular.
It was also shown that the cessation of mass transfer results in
changing of the structure of the accretion disk: the second arm
of spiral shock is appeared as well as a new dense formation
(blob), the latter moving through the disk with variable
velocity. The blob doesn't smear out under the action of
dissipation but is sustained by interaction with arms of spiral
shock for practically all the lifetime of the disk.

We also analyze the dependence of disk's lifetime on the value
of viscosity. For the value of parameter $\alpha$ that is
typical for observing accretion disk ($\alpha\sim0.01$) the
lifetime of residual disk is found to be equal to 50 orbital
periods.

\end{abstract}

\section{Introduction}

Semidetached binaries are known to be interacting binary stars
in which one component fills its Roche lobe and looses mass
through the vicinity of inner Lagrangian point $L_1$. Earlier we
developed the 3D gasdynamical model for simulation of the mass
transfer in semidetached binaries (see Bisikalo et al.
1997$^{\cite{dima97}}$, 1998b$^{\cite{dima98b}}$,
1998c$^{\cite{dima98c}}$). When applying to binaries with
constant mass transfer rate this model allows to determine the
main features of flow structure (see Bisikalo et al.
1997$^{\cite{dima97}}$, 1998b$^{\cite{dima98b}}$,
1998c$^{\cite{dima98c}}$, 1999$^{\cite{dima99}}$,
2000$^{\cite{dima2000}}$). In particular, the stream interaction
is shown to be shock-free for steady-state regime of the flow
and the model of `hot line' was introduced for binaries with
constant rate of mass transfer instead traditional `hot spot'
model.  Successful application of the `hot line' model for
analysis of light curves in cataclysmic variables (Bisikalo et
al.  1998a$^{\cite{dima98a}}$; Khruzina et al.
2001$^{\cite{tanya2001}}$) allows us to conclude that the model
appears to be adequate for considered systems.

On the other hand, some observations show that for a large
number of semidetached binaries there exist both epochs of
constant rate of mass transfer and epochs in which the rate of
mass transfer changes considerably (see, e.g., Bath et al.
1974$^{\cite{bath74}}$; Wood 1977$^{\cite{wood77}}$; Bath \&
Pringle 1982$^{\cite{bath81}}$; Bath \& van Paradijs
1983$^{\cite{bath83}}$; Gilliland 1985$^{\cite{gilliland85}}$;
Ritter 1988$^{\cite{ritter88}}$; Murray, Warner \&
Wickramasinghe 2000$^{\cite{murray2000}}$; Schreiber, G\"ansicke
\& Hessman 2000$^{\cite{schreiber2000}}$). Evolutionary
scenarios also predict the possibility of the contraction of the
mass-losing star. In this case the star doesn't fill the Roche
lobe anymore and mass transfer will be terminated.

The purpose of this paper is to investigate the flow structure
in a semidetached binary after the mass transfer termination. On
the first stage we have conducted 3D gasdynamical simulation of
mass transfer for the case of constant rate of mass transfer up
to the steady-state solution. Then we adopt that matter outflow
is ceased and consider the structure of residual disk. It is
clear that the evolution of accretion disk is controlled by
physical processes responsible for the redistribution of
the angular momentum in the disk. To investigate the influence
of viscosity we conduct 3 runs for various values of viscosity,
these ones corresponding to following values of parameter
$\alpha$ (in terms of the $\alpha$-disk):
$\alpha\sim0.08\div0.1$, $\alpha\sim0.04\div0.06$, and
$\alpha\sim0.01\div0.02$.

\section{\protect\raggedright Statement of the problem}

Let us consider a semidetached binary system and adopt that
accretor has the mass $M_1$, the mass-donating star has the mass
$M_2$, the separation of the binary system is $A$, and angular
velocity of orbital rotation is $\Omega$. The flow of matter in
this system can be described by Euler equations with equation of
state for ideal gas $P=(\gamma-1)\rho\varepsilon$, where $P$ --
pressure, $\rho$ -- density, $\varepsilon$ -- specific internal
energy, $\gamma$ -- adiabatic index. To mimic the radiative loss
of energy we adopt the value of $\gamma$ close to 1:
$\gamma=1.01$, which corresponds to the near-isothermic case
(Sawada, Matsuda \& Hachisu 1986$^{\cite{spiral1}}$; Molteni,
Belvedere \& Lanzafame 1991$^{\cite{diego91}}$; Bisikalo et al.
1997$^{\cite{dima97}}$).

To obtain the numerical solution of the system of equations we
used the Roe--Osher TVD scheme of a high approximation order
(Roe 1986$^{\cite{roe86}}$; Chakravarthy \& Osher
1985$^{\cite{osher85}}$) with Einfeldt modification (Einfeldt
1988$^{\cite{einfeldt88}}$). The original system of equations
were written in a dimensionless form. To do this, the spatial
variables were normalized to the distance between the components
$A$, the time variables were normalized to the reciprocal
angular velocity of the system ${\Omega}^{-1}$, and the density
was normalized to its value in the inner Lagrangian point $L_1$.
The gas flow was simulated over a parallelepipedon
$[\slantfrac{1}{2}A\ldots
\slantfrac{3}{2}A]\times[-\slantfrac{1}{2}A\ldots
\slantfrac{1}{2}A]\times[0\ldots \slantfrac{1}{4}A]$
(calculations were conducted only in the top half-space). The
sphere with a radius of $\slantfrac{1}{100}A$ representing the
accretor was cut out of the calculation domain.

The boundary and the initial conditions were determined as
follows:  (i) we adopted free-outflow conditions at the accretor
and at the outer boundary of the calculation domain; (ii) on the
first stage in gridpoint corresponding to $L_1$ we injected the
matter with parameters $\rho=\rho(L_1)$, $V_x=c(L_1)$,
$V_y=V_z=0$, where $c(L_1)$ is a gas speed of sound in $L_1$
point; (iii) for this stage we used rarefied background gas with
the following parameters $\rho_0=10^{-5}\cdot\rho(L_1)$,
$P_0=10^{-4}\rho(L_1)c^2(L_1)/\gamma$, ${\bmath V}_0=0$ as the
initial conditions; (iv) on the second stage when steady-state
regime is reached at the moment of time $t=t_0$ we decreased the
density of the injected matter to the value $\rho=\rho_0$.

The analysis of considered problem shows that the gas dynamical
solution for the semidetached binary is defined by three
dimensionless parameters (Bisikalo et al.
1998b$^{\cite{dima98b}}$, 1999$^{\cite{dima99}}$; Lubow \&
Shu 1975$^{\cite{lubowshu75}}$): the mass ratio $q=M_2/M_1$,
the Lubow-Shu parameter $\epsilon=c(L_1)/A\Omega$ (Lubow \& Shu
1975$^{\cite{lubowshu75}}$), and the adiabatic index $\gamma$.
The value of the adiabatic index was discussed above and we used
the value $\gamma=1.01$. Analysis of our previous results
(Bisikalo et al. 1998b$^{\cite{dima98b}}$,
1999$^{\cite{dima99}}$) shows that the main characteristic
features of 3D gas dynamical flow structure are qualitatively
the same in wide range of parameters $q$ and $\epsilon$.
Therefore for the model simulation we chose them as follows:
$q=1$, $\epsilon=\slantfrac{1}{10}$.

To evaluate the influence of the viscosity on the solution,
several runs with different spatial resolution were conducted.
The Euler equations do not include physical viscosity so we
varied the numerical viscosity by virtue of changing of
computational grid.  Three grids were chosen for simulation:
$31\times31\times17$, $61\times61\times17$, $91\times91\times25$
(hereinafter runs "A", "B", and "C", correspondingly). In terms
of the $\alpha$-disk the numerical viscosity for runs "A", "B",
"C" approximately corresponds $\alpha\sim0.08\div0.1$,
$\alpha\sim0.04\div0.06$, and $\alpha\sim0.01\div0.02$.

\section{\protect\raggedright Flow structure after the mass
transfer termination}

Prior to simulation of the flow structure with `turned-off' mass
transfer the near-steady-state solutions for the case of
constant non-zero rate of mass transfer were obtained (Bisikalo
et al. 2000$^{\cite{dima2000}}$) and used as initial conditions.
At the moment of time $t=t_0$ the rate of mass transfer was
decreased in five order of magnitude, which is correspond to the
cessation of mass transfer.

Calculations of runs "A" and "B" were lasted until the density
of gas becomes less than the background density
$\rho_0=10^{-5}\rho(L_1)$. This corresponds to the full
vanishing of matter due to accretion and outflow through
outer boundary. The durations of these two runs correspond to
$5P_{orb}$ and $12P_{orb}$ after the moment of time $t_0$. Run
"C" has the best resolution and minimal numerical
viscosity. Nevertheless this run was conducted to $10P_{orb}$
only, since this run is very computer-time-consuming while the
time when disk vanishes can be estimated as $\sim50P_{orb}$. We
extrapolated the results of calculation of run "C" for
$t>10P_{orb}$.

\begin{figure}[t]
\centerline{\hbox{\psfig{figure=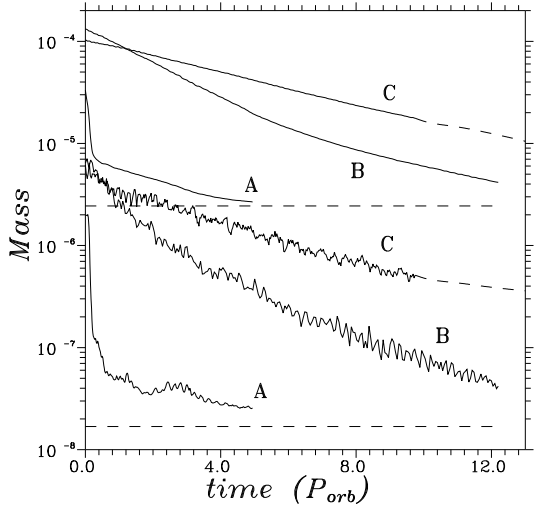,width=7cm}}}
\caption{\footnotesize Time evolution of the total mass of the
calculation domain (top part of the figure) and the mass of the
disk (bottom part of the figure) for "A", "B", and "C" runs.
The mass of uniform matter with background density $\rho=\rho_0$
for the whole computation domain and for the disk are shown by
dashed lines.} \end{figure}

Before investigating of the structure of gaseous flows in binary
system let us estimate the lifetime of the residual disk after
the termination of mass transfer for different values of
viscosity.  The lifetime of the residual disk can be evaluated
from the Fig.~1 where the total mass of the calculation
domain (top part of the figure) and the mass of the disk (bottom
part of the figure) versus time is presented. The lower limit of
the density of gas is the background density $\rho_0$, therefore
the mass of the uniform matter with $\rho=\rho_0$ calculated for
the whole computation domain and for the disk are asymptotic
lines for graphs $M(t)$.  These lines are shown in the Fig.~1 by
dashed lines.

The analysis of the data presented in the Fig.~1 shows (and
there is no surprise now) that the lifetime of the residual disk
is increased when the numerical viscosity is decreased: for run
"A" ($\alpha\sim0.1$) $\tau_{disk}=5P_{orb}$; for run "B"
($\alpha\sim0.05$) $\tau_{disk}=12P_{orb}$; and for run "C"
($\alpha\sim0.05$) the extrapolation of the result of simulation
shows that the lifetime of the residual disk exceeds
$50P_{orb}$.  It is pertinent to note that the value
$\alpha\sim0.01$ is typical for observable accretion disks (see,
e.g., Lynden-Bell \& Pringle 1974$^{\cite{lyndenbell74}}$;
Meyer-Hofmeister \& Ritter 1993$^{\cite{meyer93}}$; Armitage
\& Livio 1996$^{\cite{armi96}}$; Tout 1996$^{\cite{tout96}}$).
Consequently, we can expect the lifetime of residual disk after
the termination of mass is of order $50P_{orb}$ (near a week)
for typical dwarf novae.

\begin{figure*}[p]
\centerline{\hbox{\psfig{figure=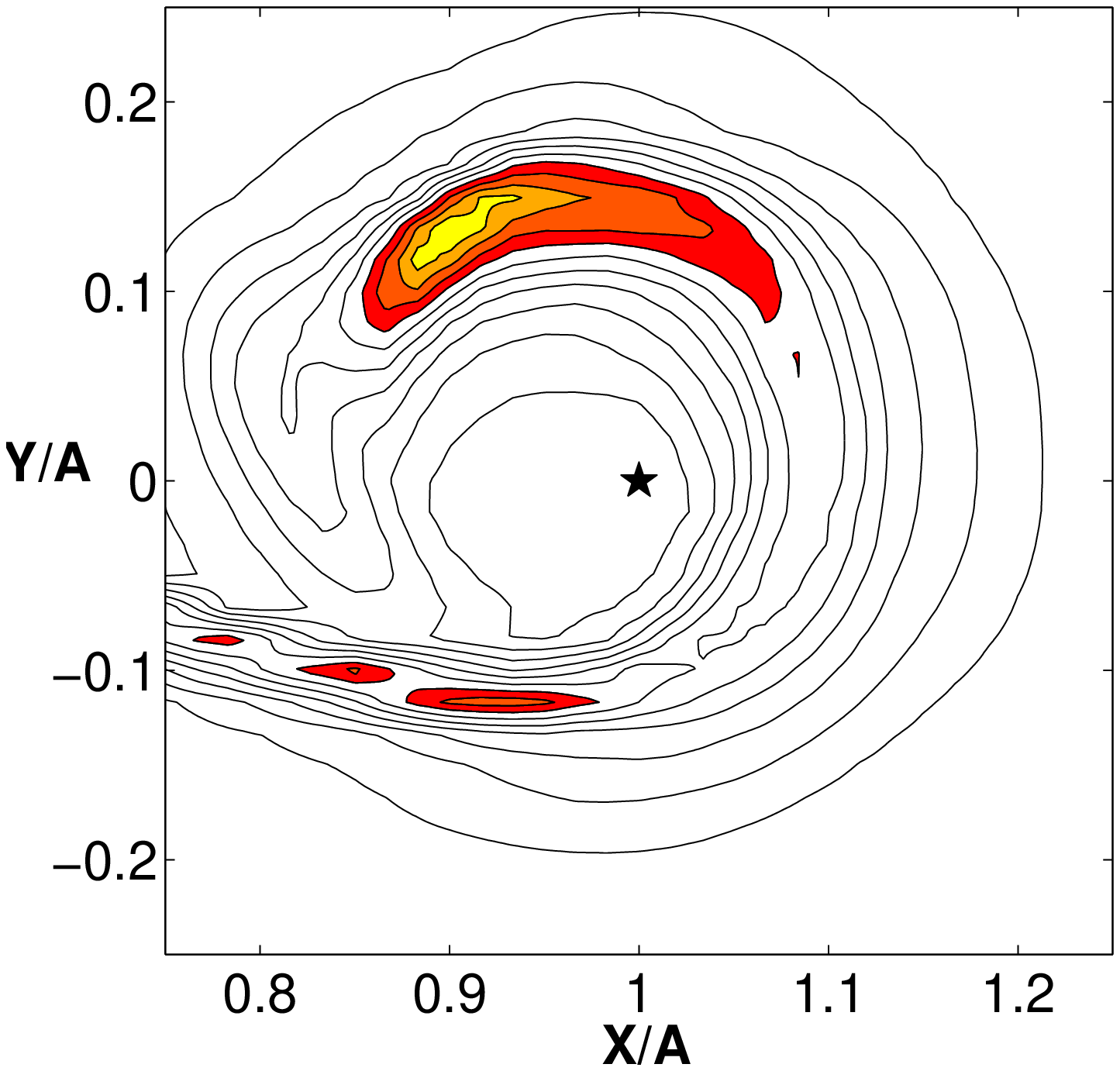,width=6.5cm}}
\hbox{\psfig{figure=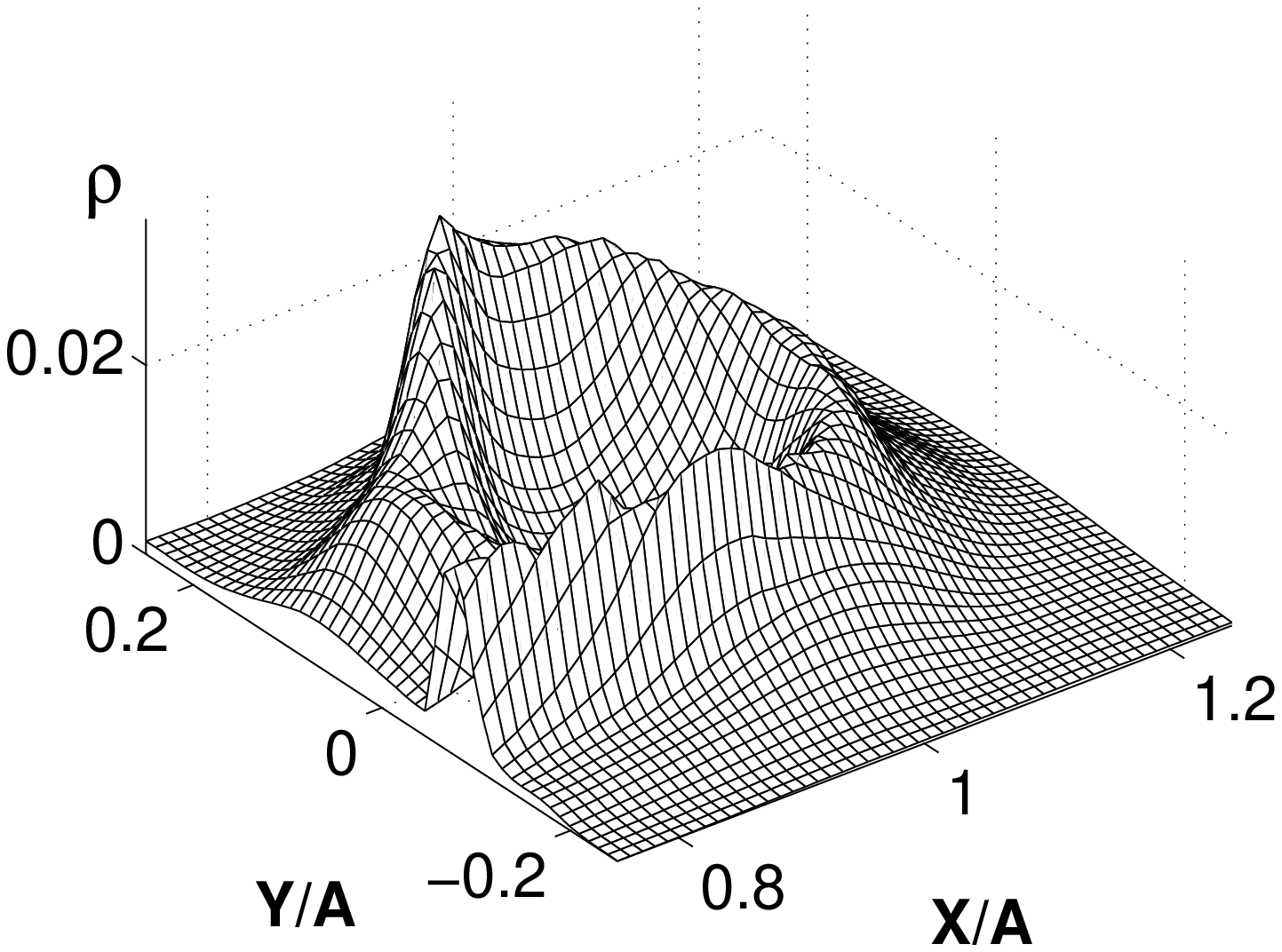,width=8.5cm}}}
\centerline{\hbox{\psfig{figure=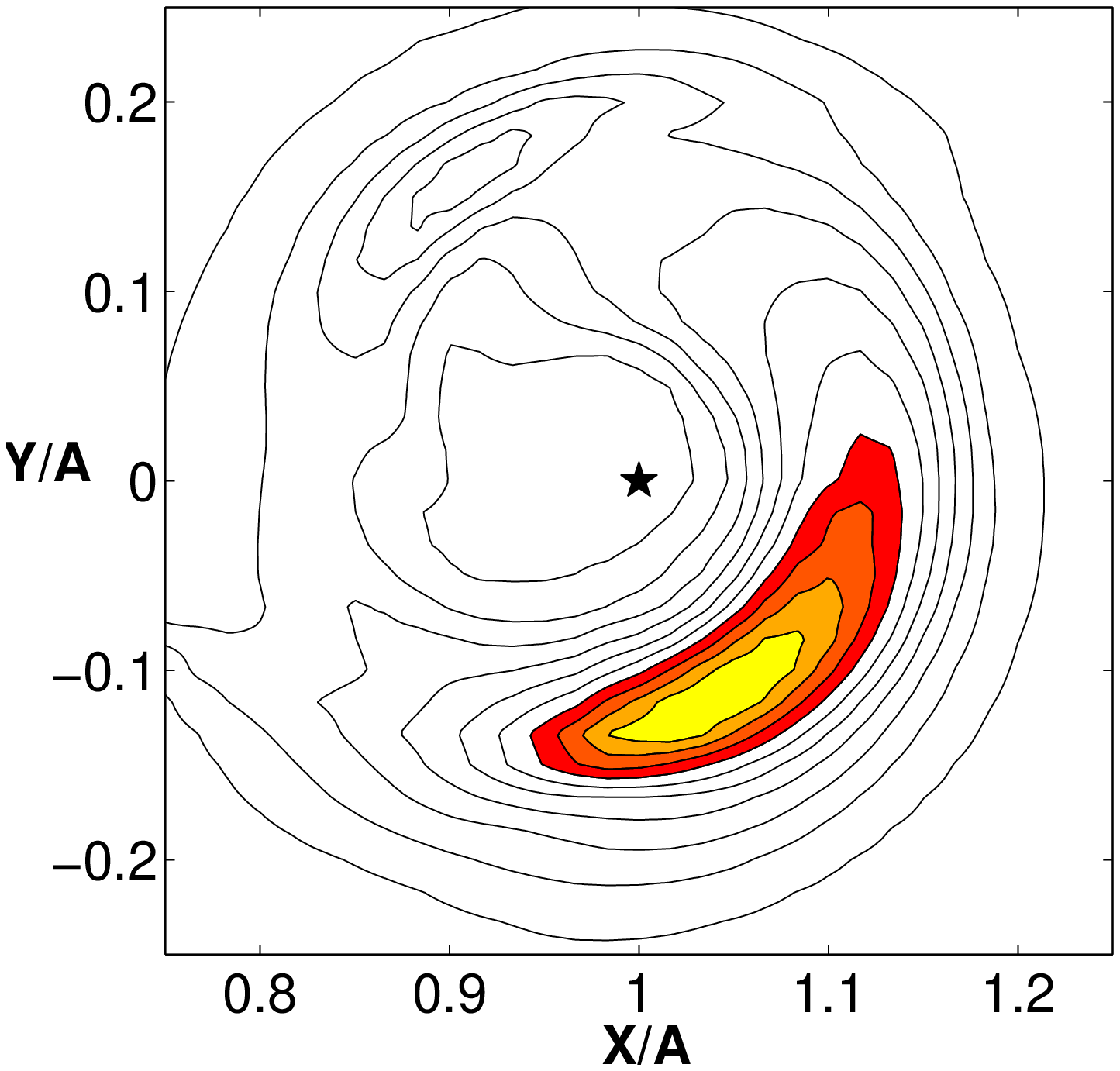,width=6.5cm}}
\hbox{\psfig{figure=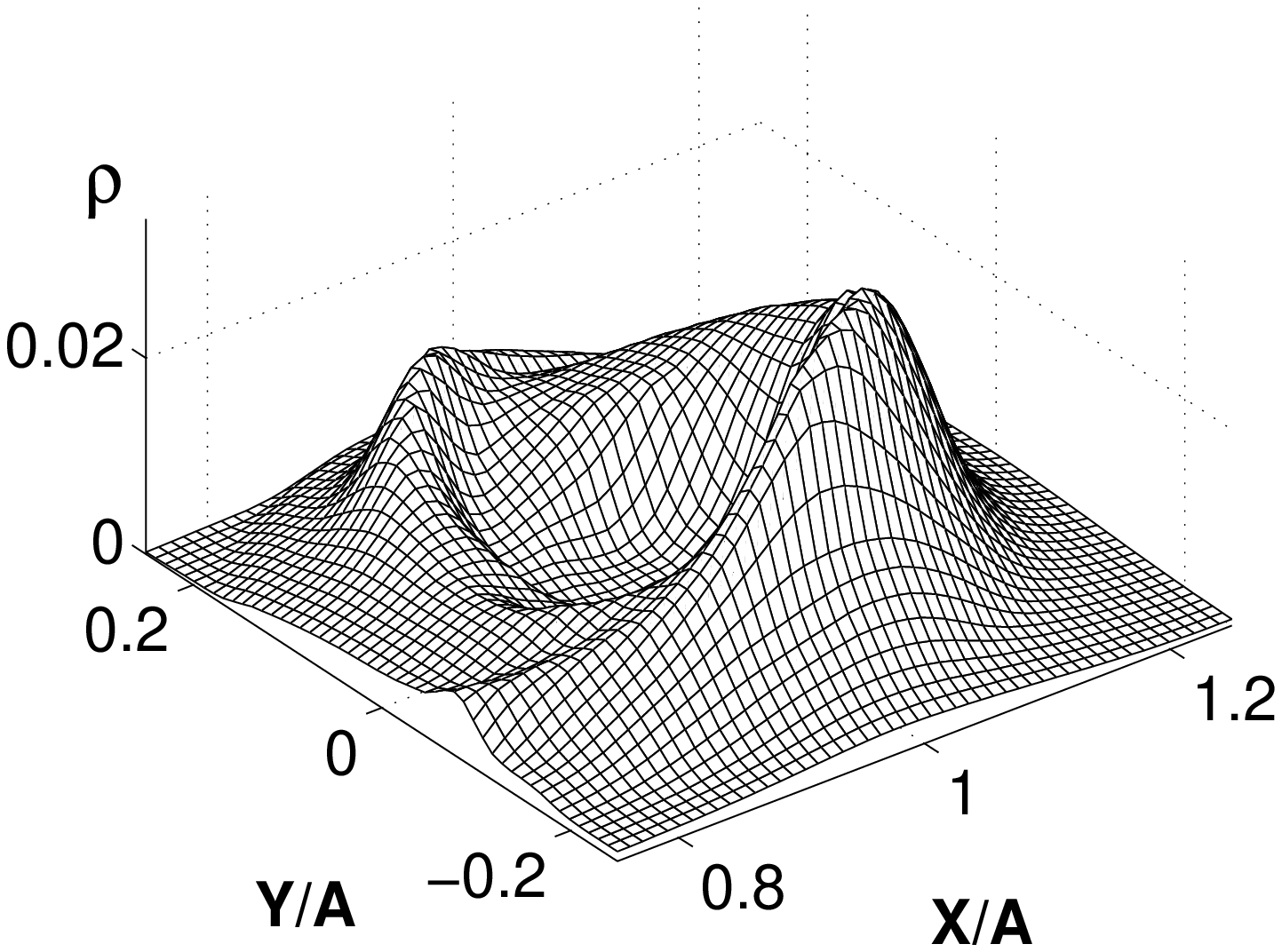,width=8.5cm}}}
\centerline{\hbox{\psfig{figure=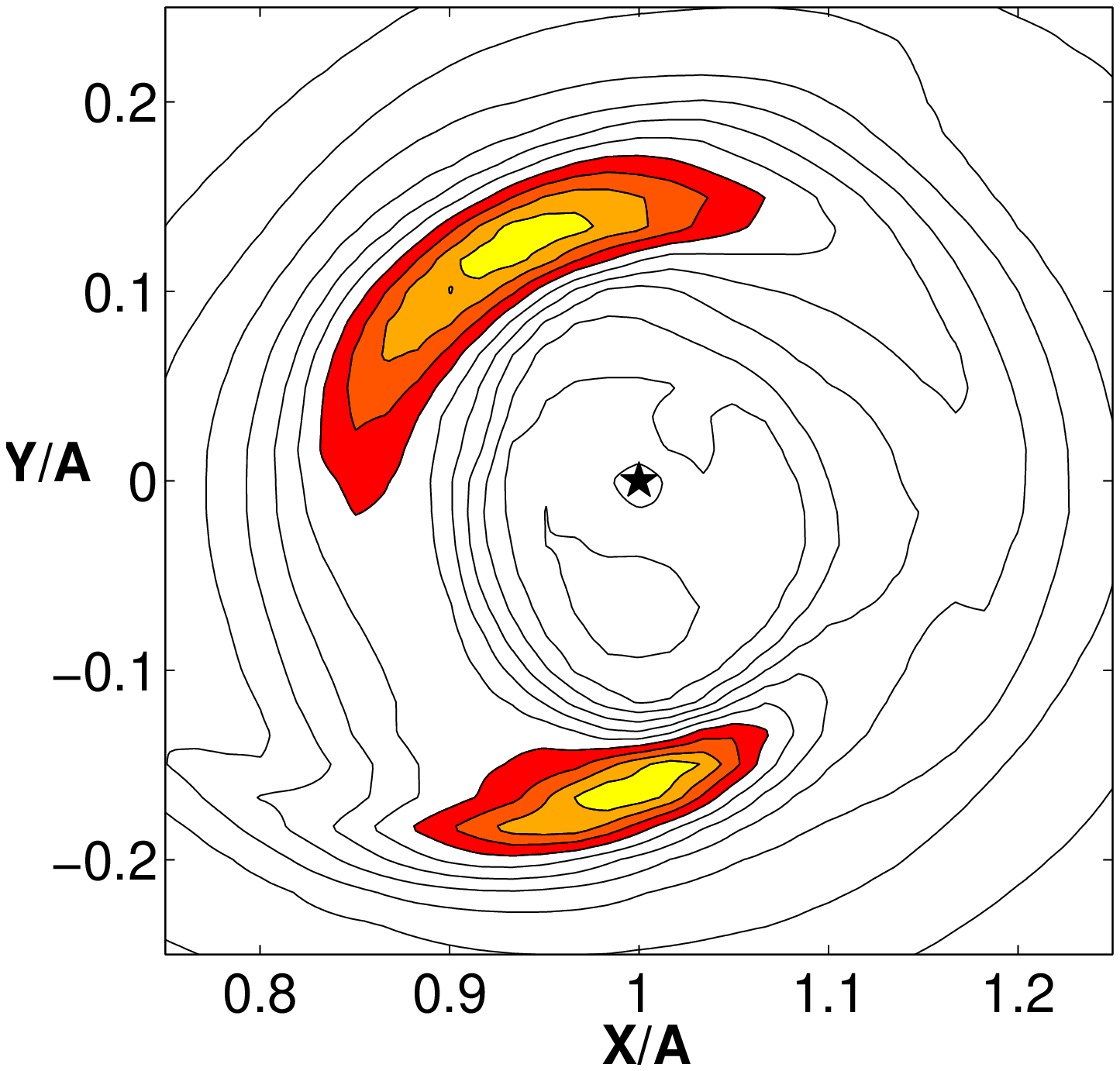,width=6.5cm}}
\hbox{\psfig{figure=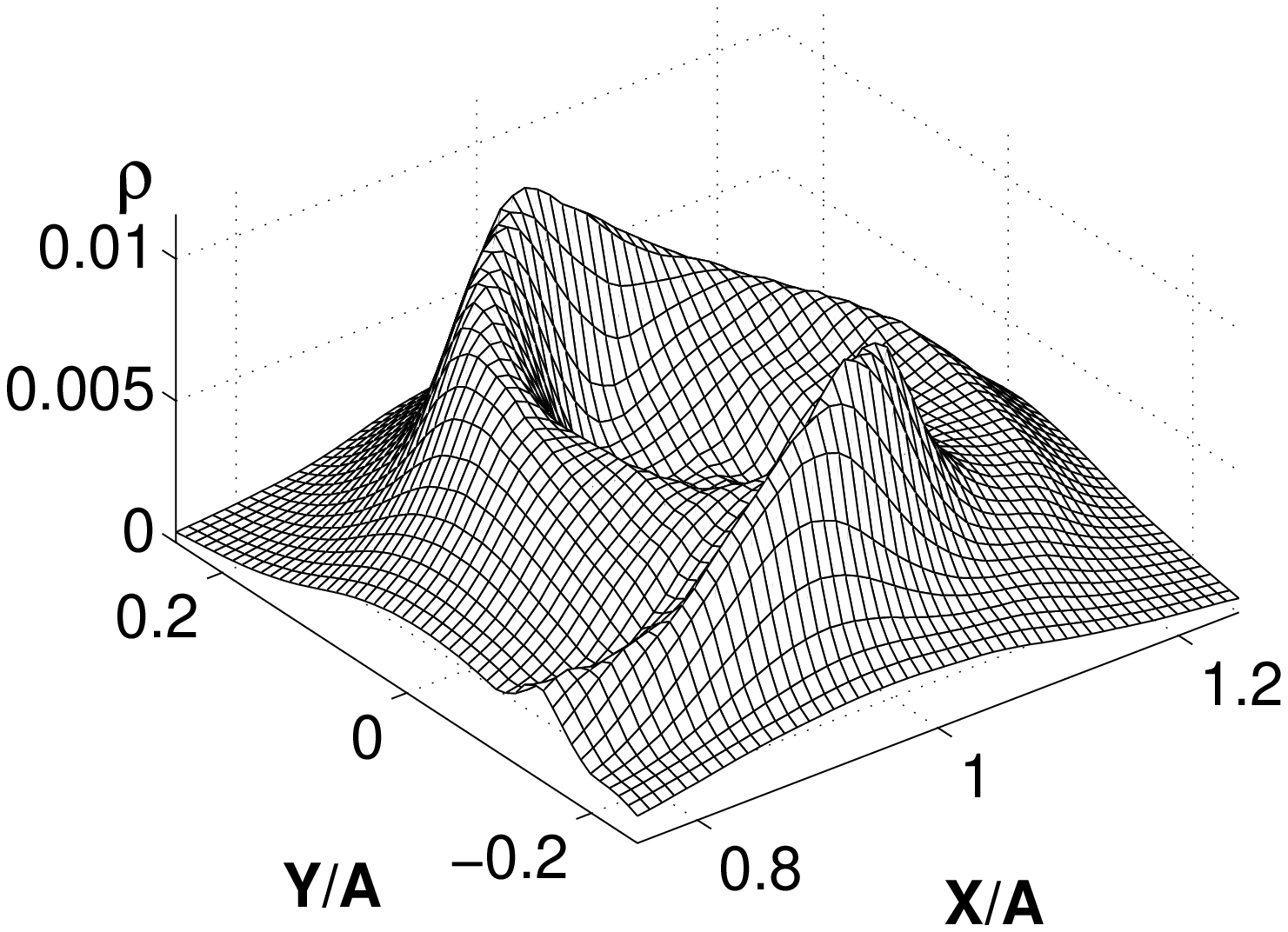,width=8.5cm}}}
\caption{\footnotesize Isolines (left panels) and bird-eye vies
(right panels) of density in the equatorial plane. Three rows of
panels correspond to the following moments of time:
$t=t_0+0.08P_{orb}$, $t=t_0+0.16P_{orb}$, $t=t_0+0.9P_{orb}$,
where $t_0$ is the moment of the mass transfer termination.}
\end{figure*}

Now let us consider the  evolution of the residual accretion
disk in time. Since the results of runs "A", "B", "C" are
qualitatively similar, we will focus on the run "B" and analyze
it in detail.  The choice of run "B" is called forth by the
following reasons:  firstly, the spatial resolution of this run
is well enough to catch all the details of the disk structure,
and, secondly, the lifetime of residual accretion disk is
sufficiently small ($12P_{orb}$) so we can study its fate up to
full vanishing.  The Figure~2 presents isolines and bird-eye
views of density in the equatorial plane for different moments
of time. the upper panels correspond to the moment of time
$t=t_0+0.08P_{orb}$, middle panels -- to $t=t_0+0.16P_{orb}$,
and lower panels -- to $t=t_0+0.9P_{orb}$.  The analysis of
these figures shows that for initial moments of time
($t<t_0+0.08P_{orb}$) the structure of accretion disk doesn't
differ much from the structure obtained for the steady-state
solution with constant mass transfer rate (see Bisikalo et al.
1997$^{\cite{dima97}}$, 1998b$^{\cite{dima98b}}$,
1998c$^{\cite{dima98c}}$, 1999$^{\cite{dima99}}$,
2000$^{\cite{dima2000}}$). As always the stream of matter from
$L_1$ dominates. It is clearly seen that the uniform morphology
of the system stream--disk results in quasi-elliptical shape of
the disk and the absence of `hot spot' in zone of stream--disk
interaction.  At the same time the interaction of the gas of
circumbinary envelope with the stream results in the formation
of an extended shock wave located along the stream edge (`hot
line'). The Figure~2 also manifests the formation of tidally
induced spiral shock located in I and II quadrants of coordinate
plane.  Appearance of tidally induced two-armed spiral shock was
discovered in Sawada et al.  (1986a$^{\cite{spiral1}}$,
1986b$^{\cite{spiral2}}$, 1987$^{\cite{sawada87}}$). Here we see
only the one-armed spiral shock. The flow structure in the
place where the second arm should be is defined by the stream
from $L_1$ which appears to suppress the formation of another
arm of the tidally induced spiral shock.

\begin{figure}[t]
\centerline{\hbox{\psfig{figure=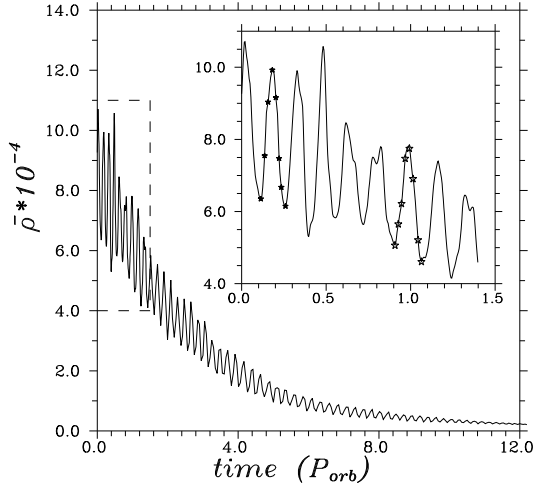,width=7cm}}}
\caption{\footnotesize Time variation of the mean density of
matter passing through a semi-plane $XZ$ ($Y=0$, $X>A$) slicing
the disk. A zoom for the mean density variation for times from
$t_0$ to $t_0+1.4P_{orb}$ is shown in the top of the figure.
Two sets of moments of time (8 moments per set, each set covers
one period of the mean density variation) are also shown.}
\end{figure}

From the panels of middle row of the Fig.~2 it is seen that after
the moment of time $t\sim t_0+0.16P_{orb}$ the stream from $L_1$
has no influence on the structure of residual accretion disk
anymore and the disk is circularized. It is also seen that the
shock wave located along the stream edge (`hot line') disappears
and the second arm of the tidally induced spiral shock located in
III and IV quadrants of coordinate plane is formed. It is
illustrated by lower row of the Fig.~2 where results of
simulations for the moment of time $t=t_0+0.16P_{orb}$ are
presented. Note that the intensity of shocks as well as the mass
of accretion disk decreased in time but the qualitative flow
structure retains. The flow structure is changed when the tidal
interaction doesn't result in the formation of spiral shocks
anymore (it corresponds to the moments of time $t_0+10P_{orb}
\div t_0+11P_{orb}$ for run "B").  Therefore, we can conclude
that the spiral shocks exists during of all lifetime of the disk
(remind that it is of order $12P_{orb}$ for run "B").

\renewcommand{\thefigure}{4a}
\begin{figure*}[p]
\centerline{\hbox{\psfig{figure=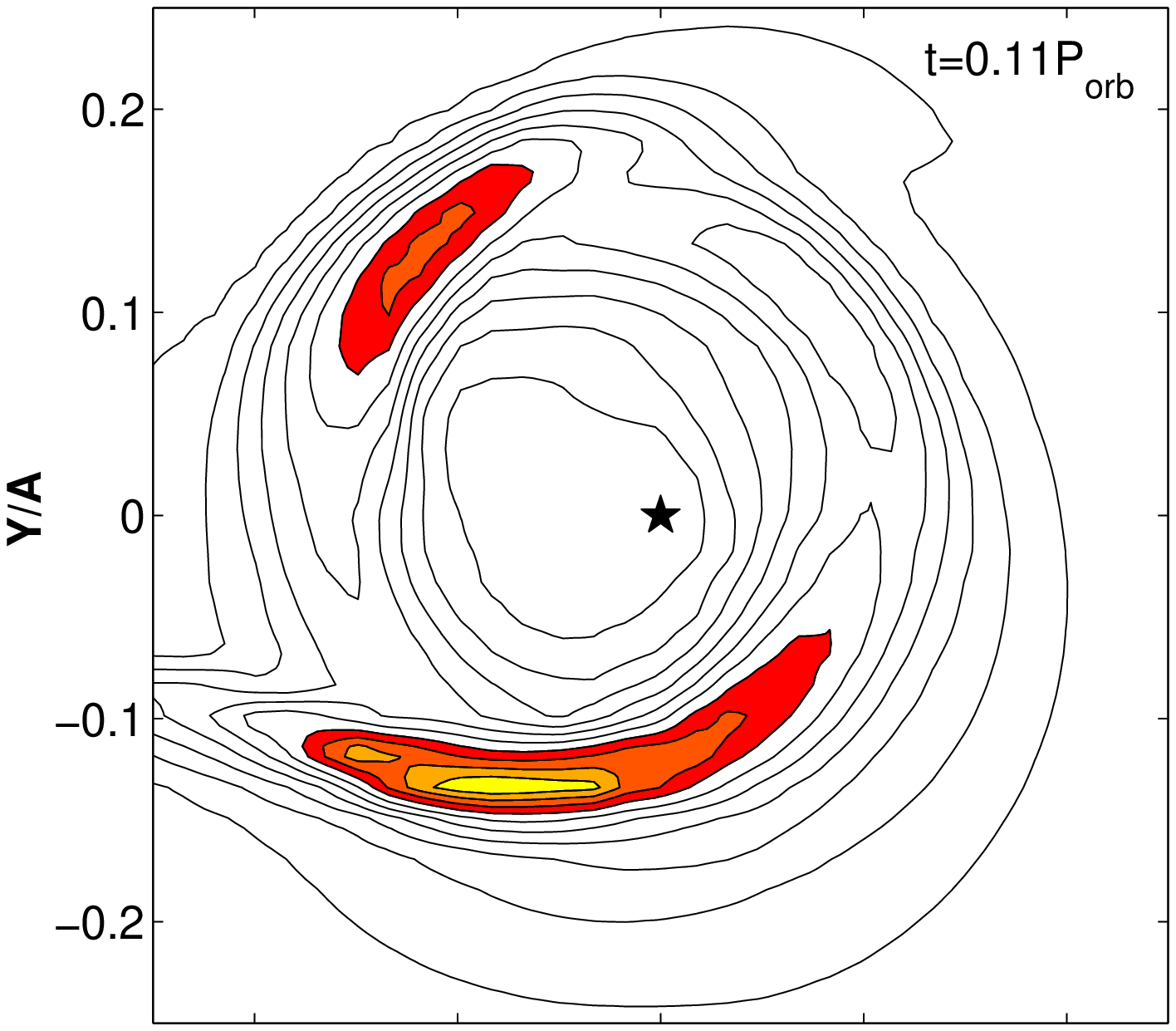,width=58mm}}
\hbox{\psfig{figure=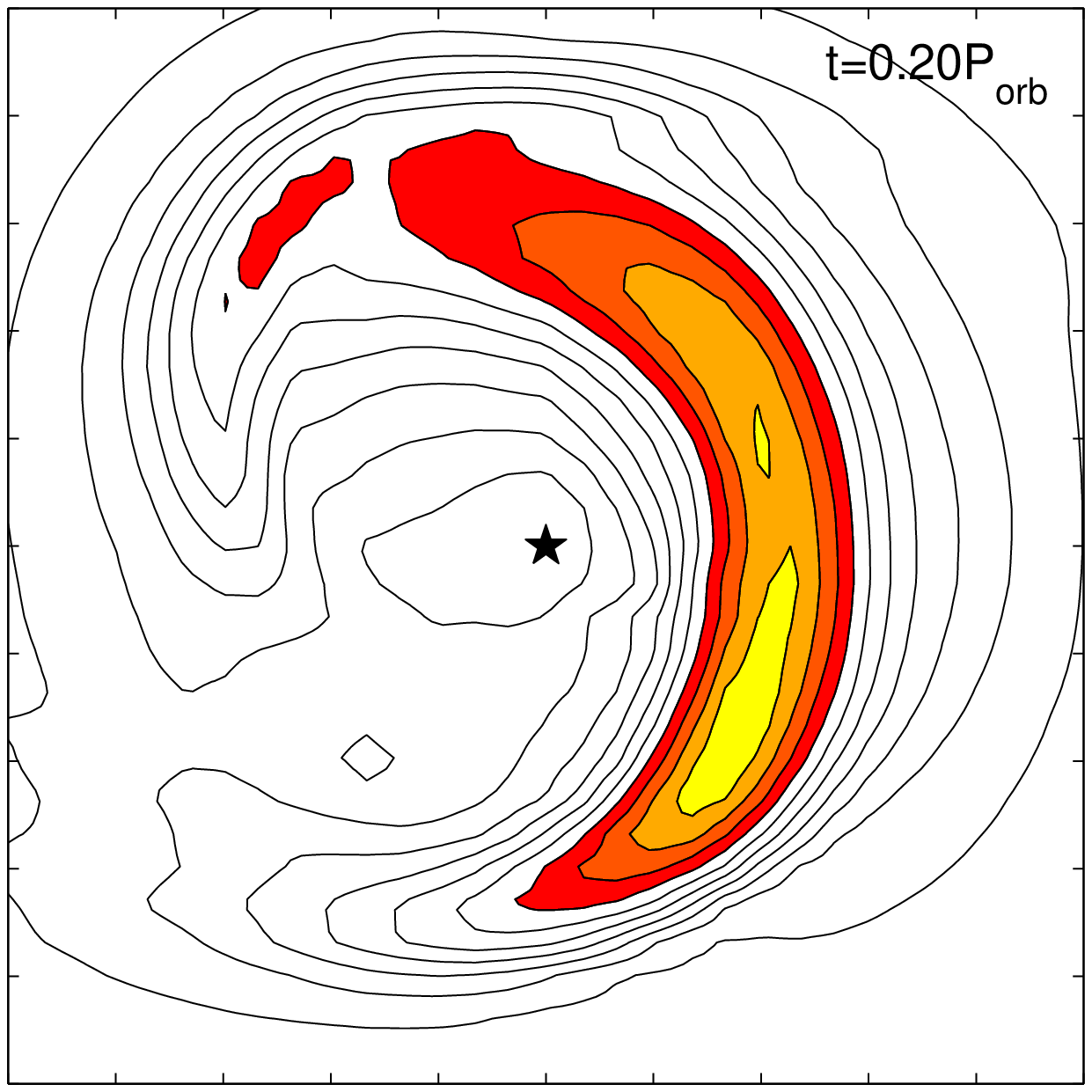,width=51.5mm}}}
\centerline{\hbox{\psfig{figure=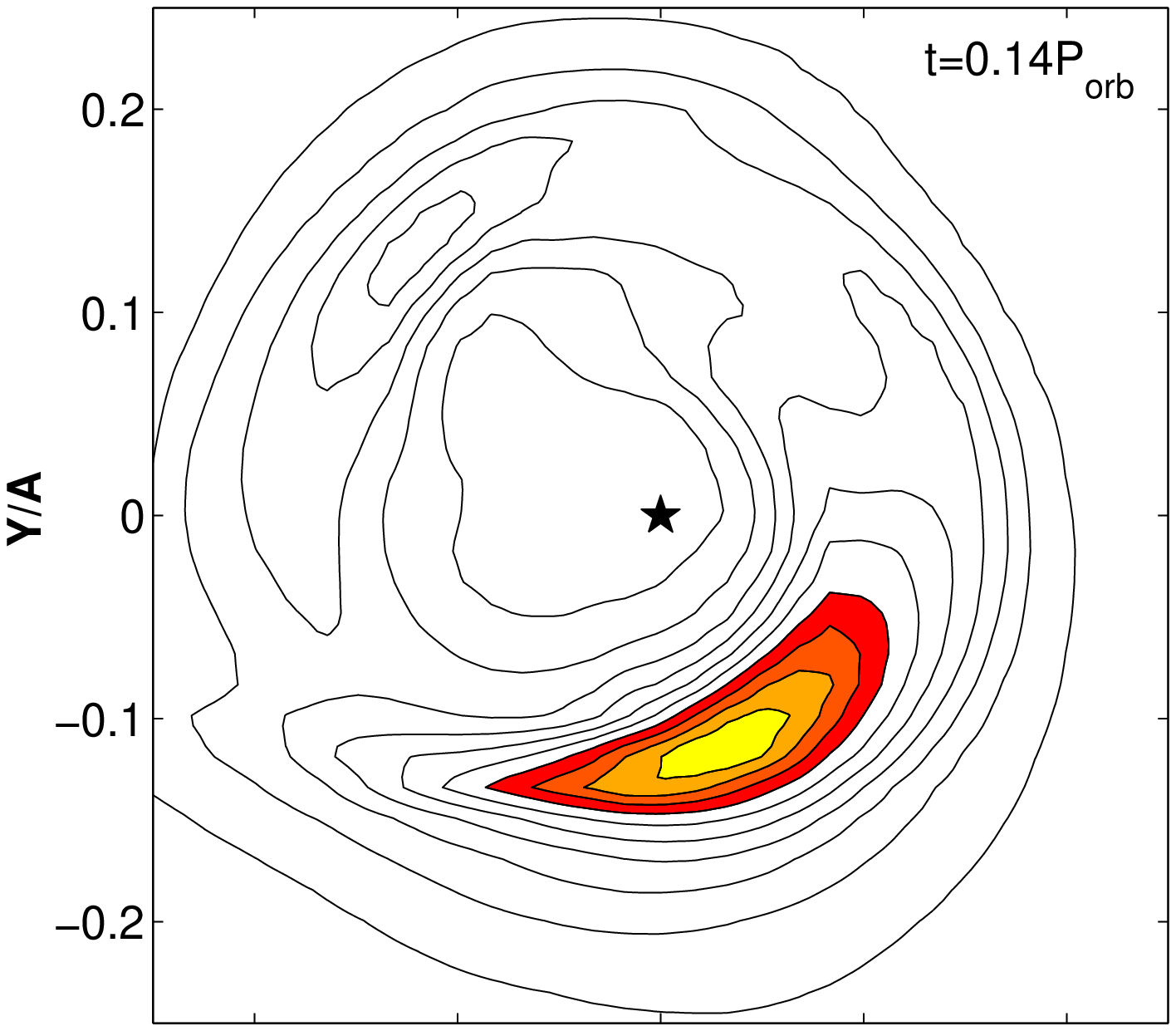,width=58mm}}
\hbox{\psfig{figure=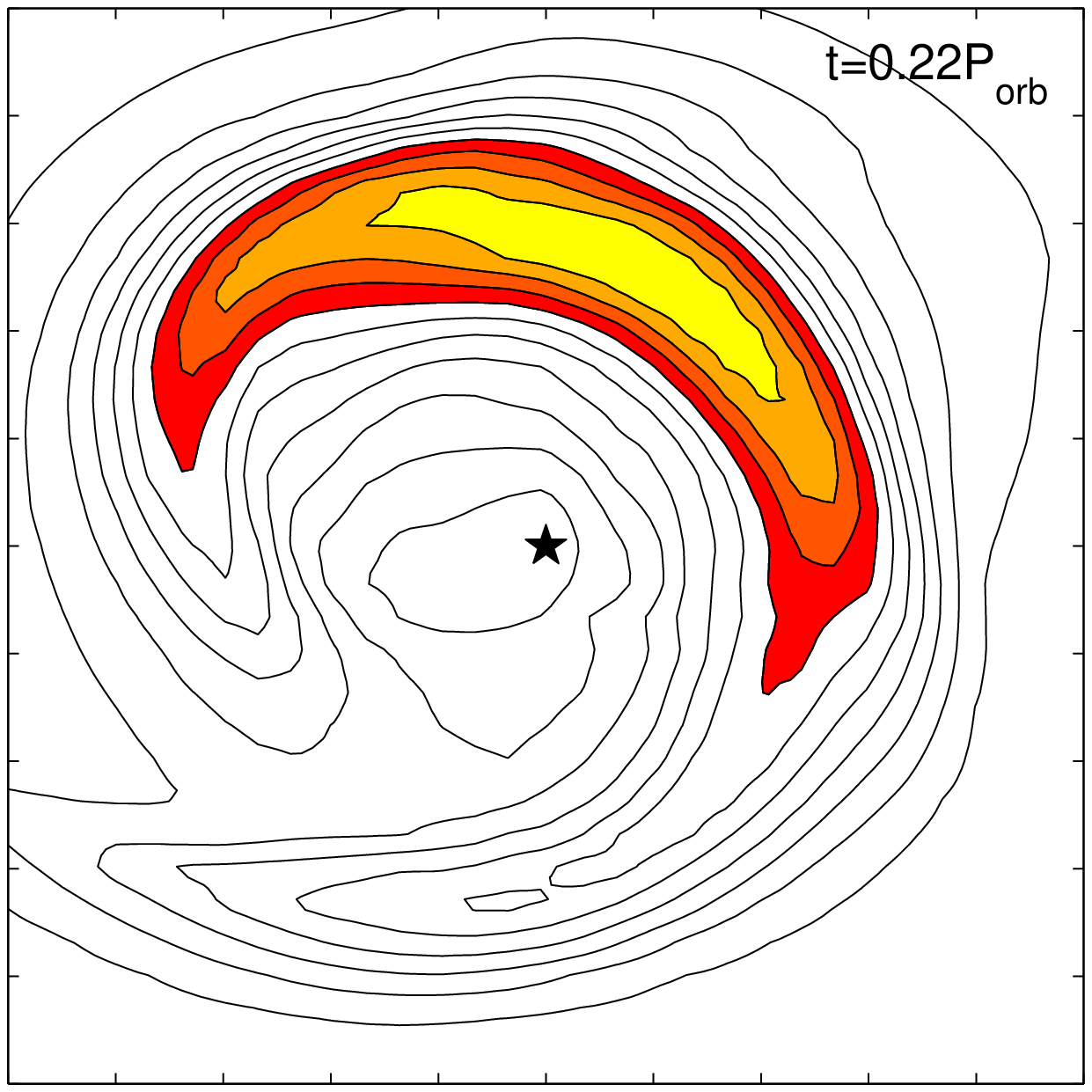,width=51.5mm}}}
\centerline{\hbox{\psfig{figure=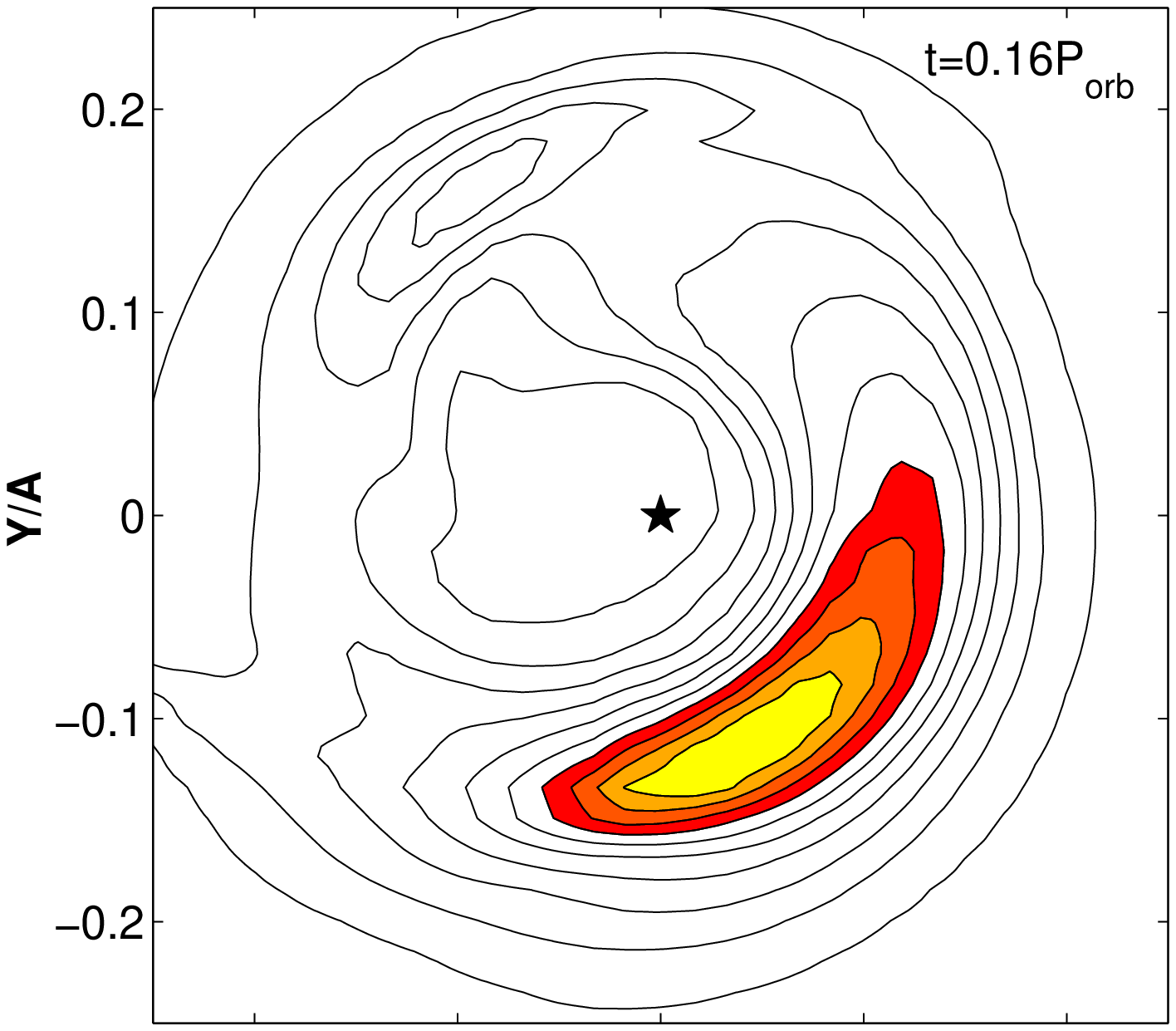,width=58mm}}
\hbox{\psfig{figure=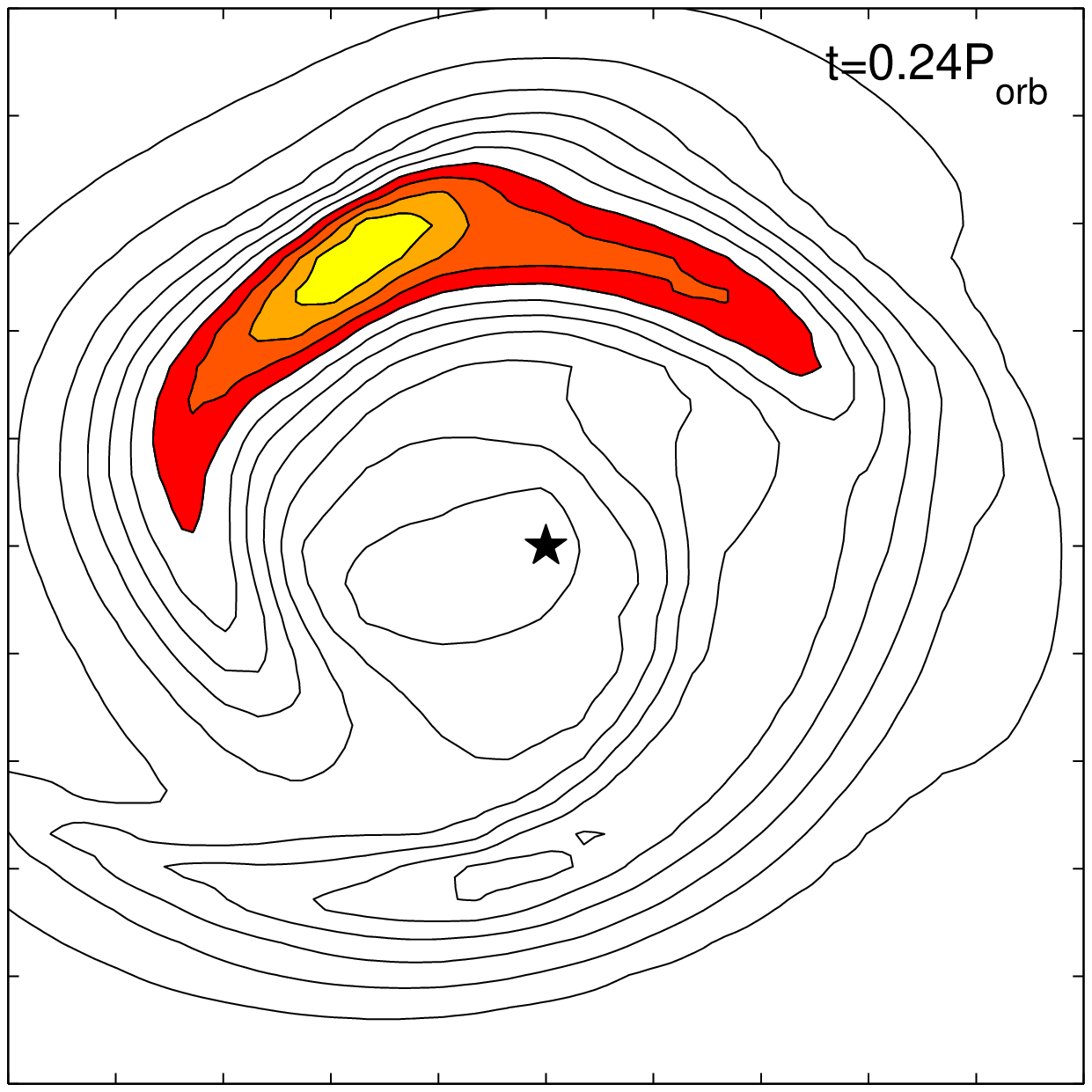,width=51.5mm}}}
\centerline{\hbox{\psfig{figure=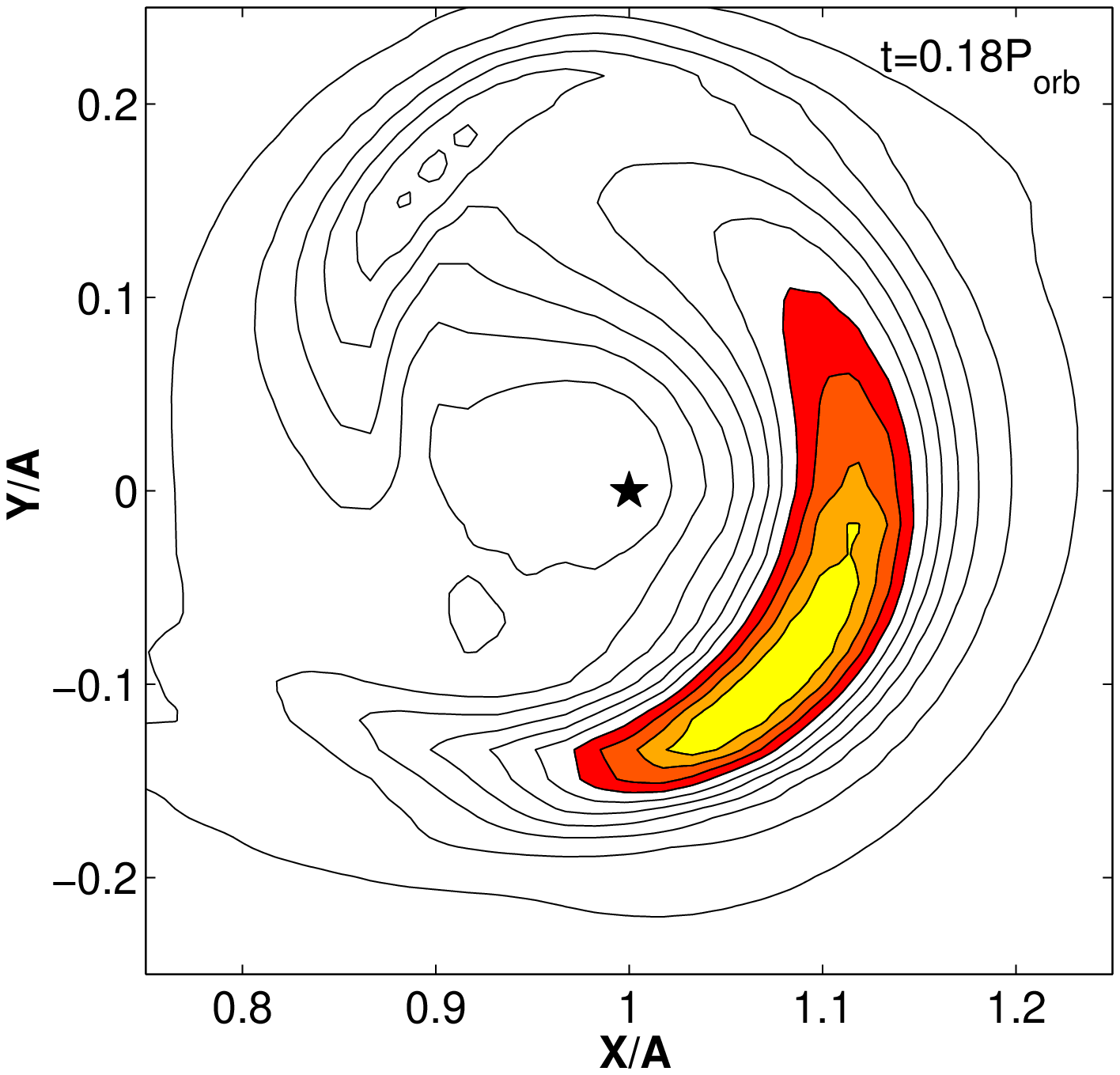,width=58mm}}
\hbox{\psfig{figure=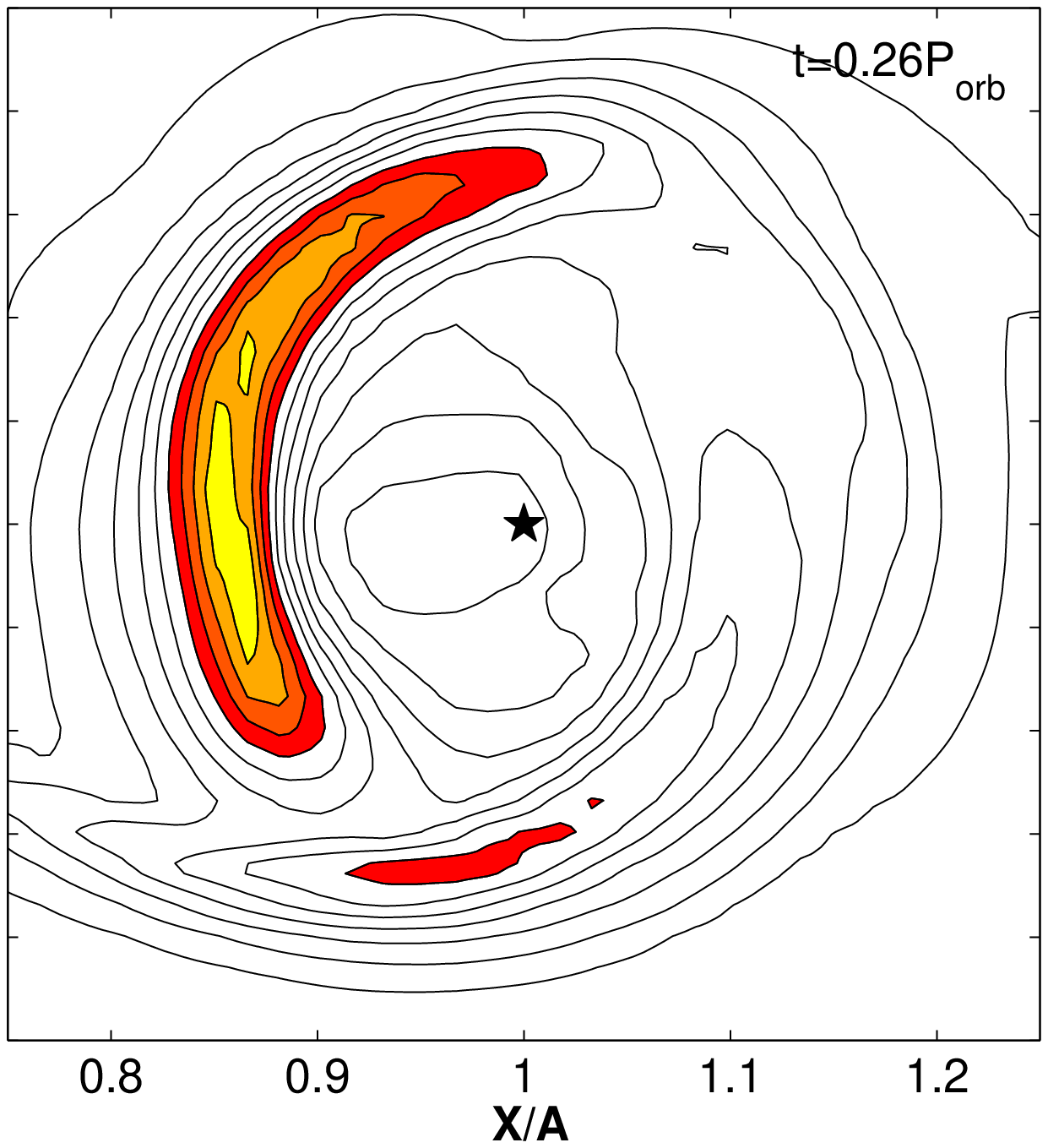,width=51.5mm}}}
\caption{\footnotesize Density distribution in the equatorial
plane for the first set of moments of time from Fig.~3.}
\end{figure*}

\renewcommand{\thefigure}{4b}
\begin{figure*}[p]
\centerline{\hbox{\psfig{figure=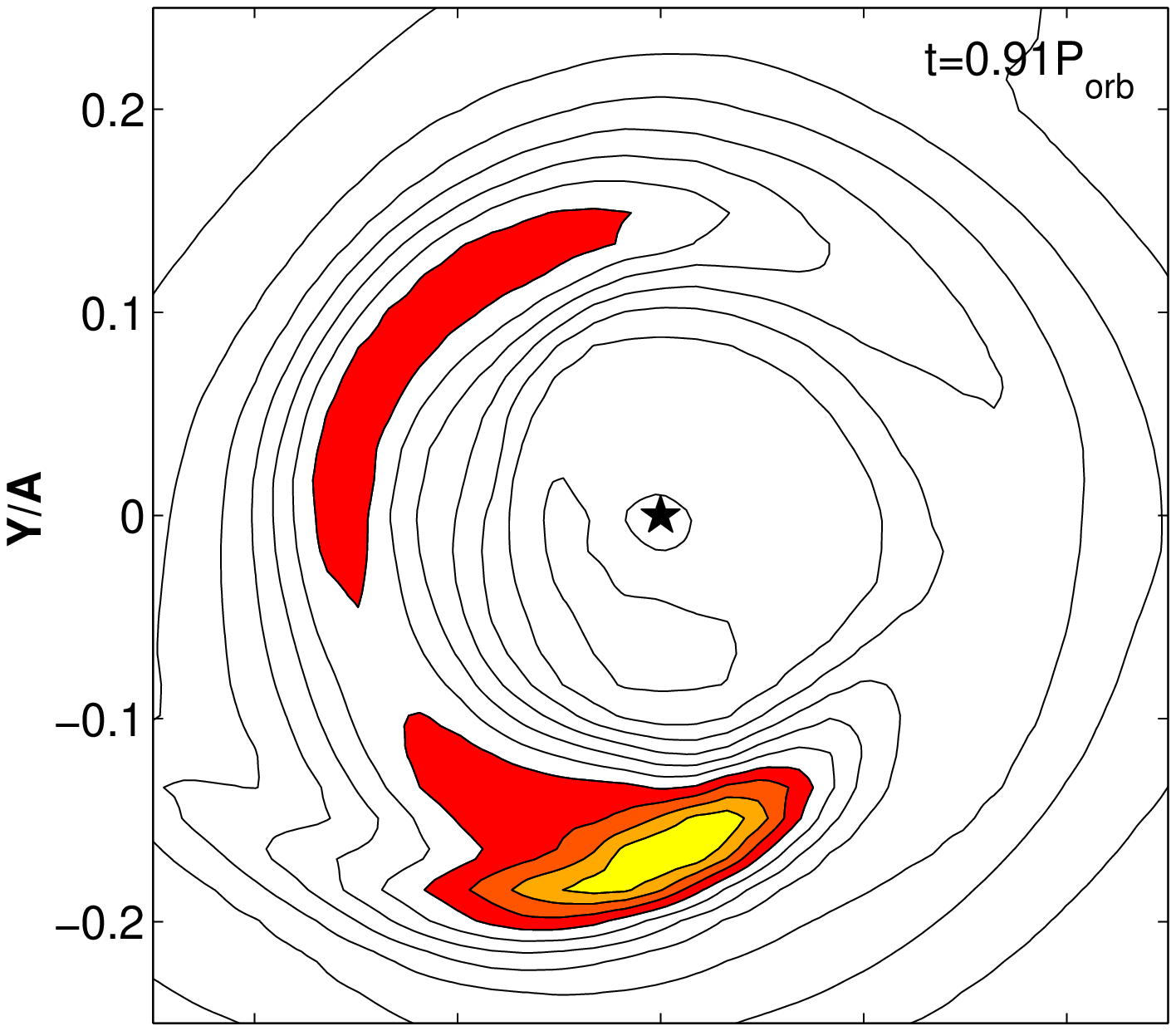,width=58mm}}
\hbox{\psfig{figure=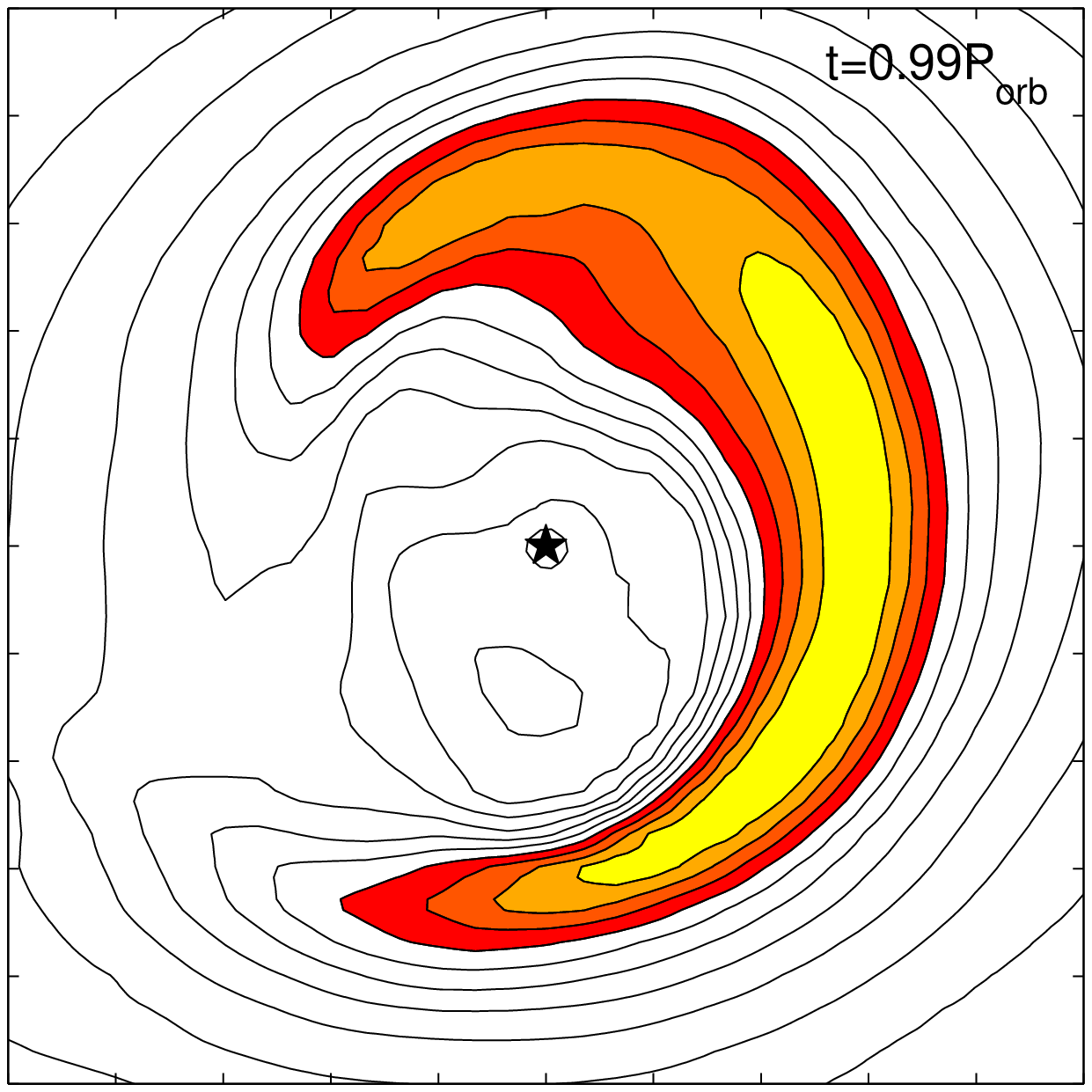,width=51.5mm}}}
\centerline{\hbox{\psfig{figure=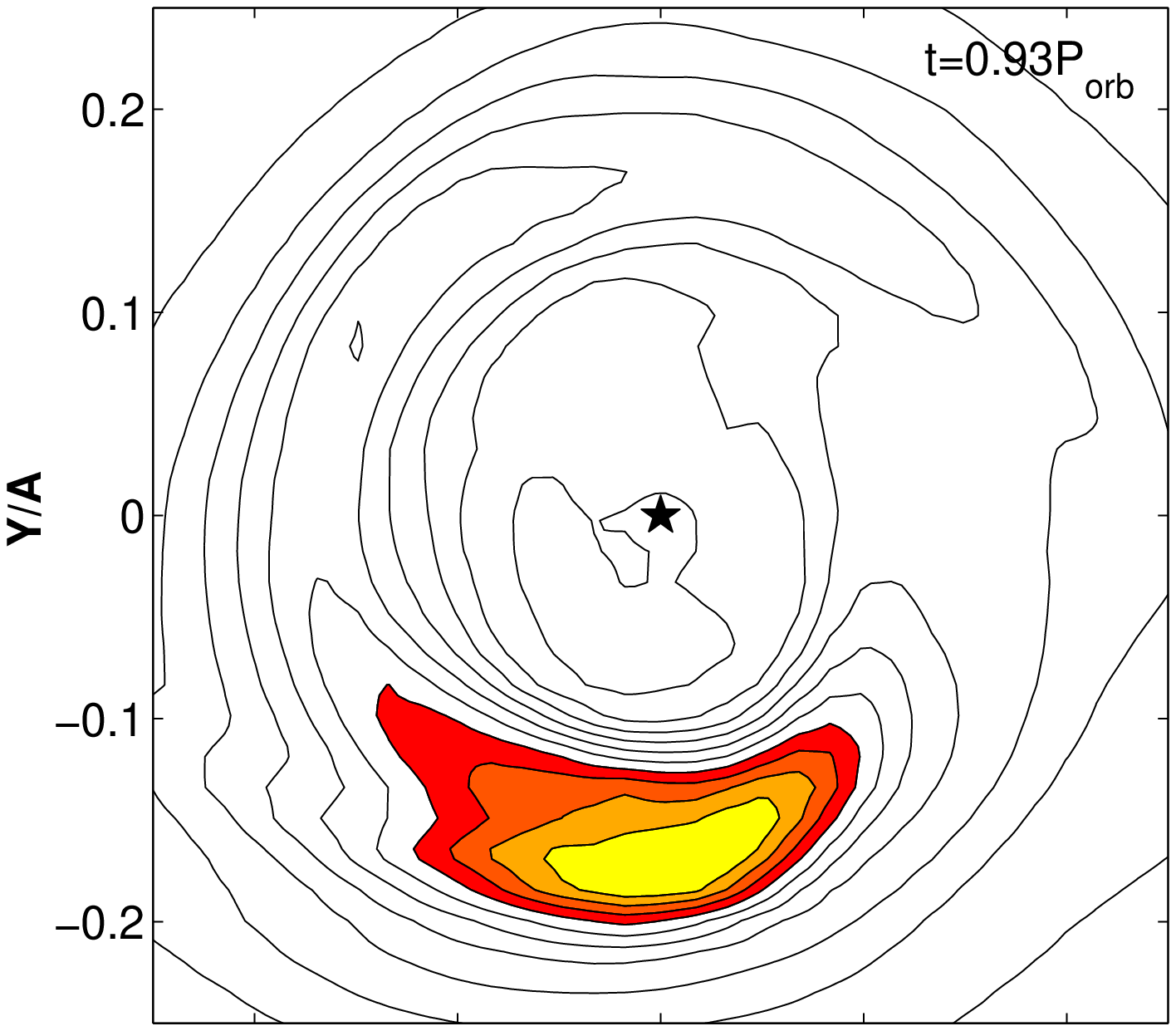,width=58mm}}
\hbox{\psfig{figure=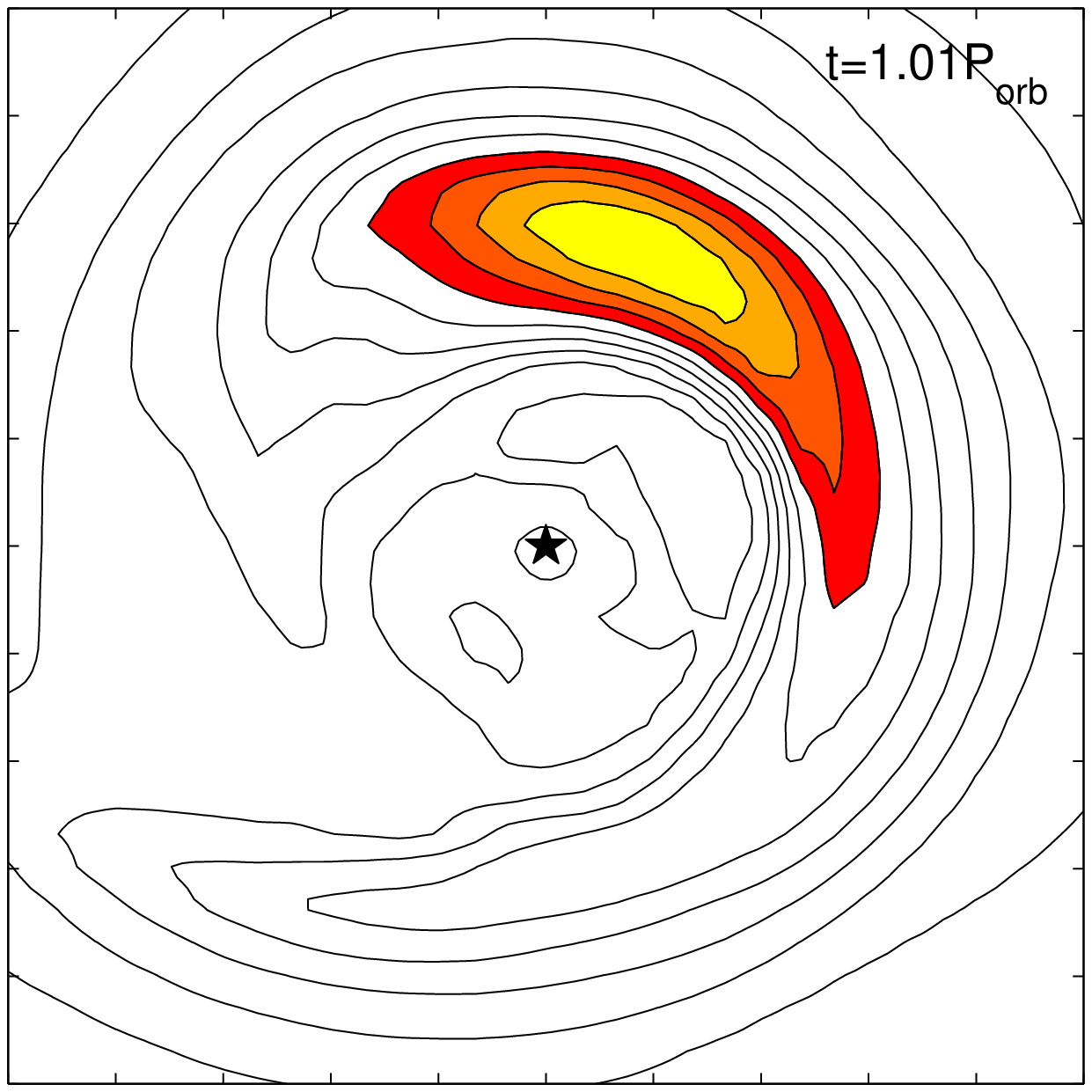,width=51.5mm}}}
\centerline{\hbox{\psfig{figure=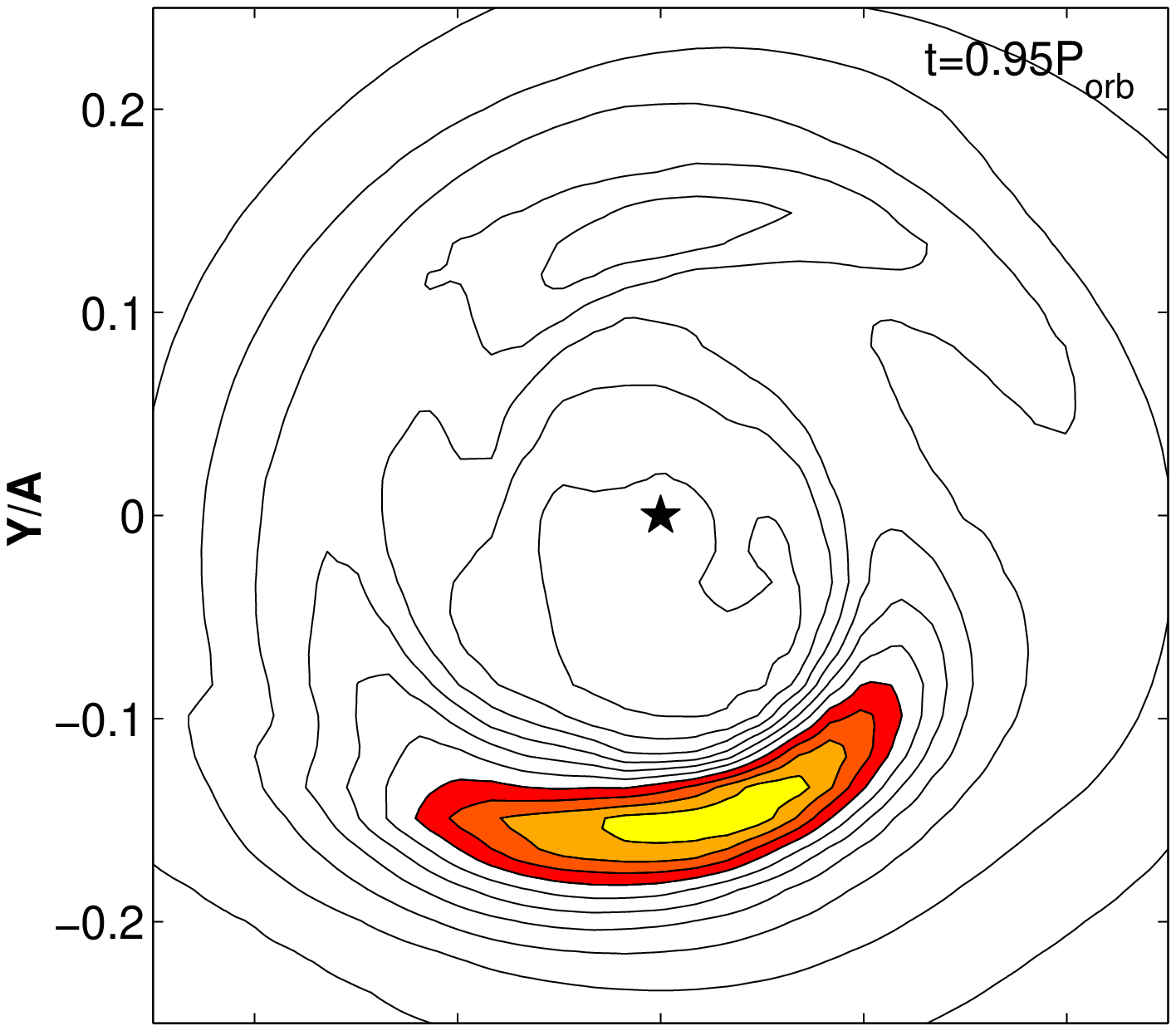,width=58mm}}
\hbox{\psfig{figure=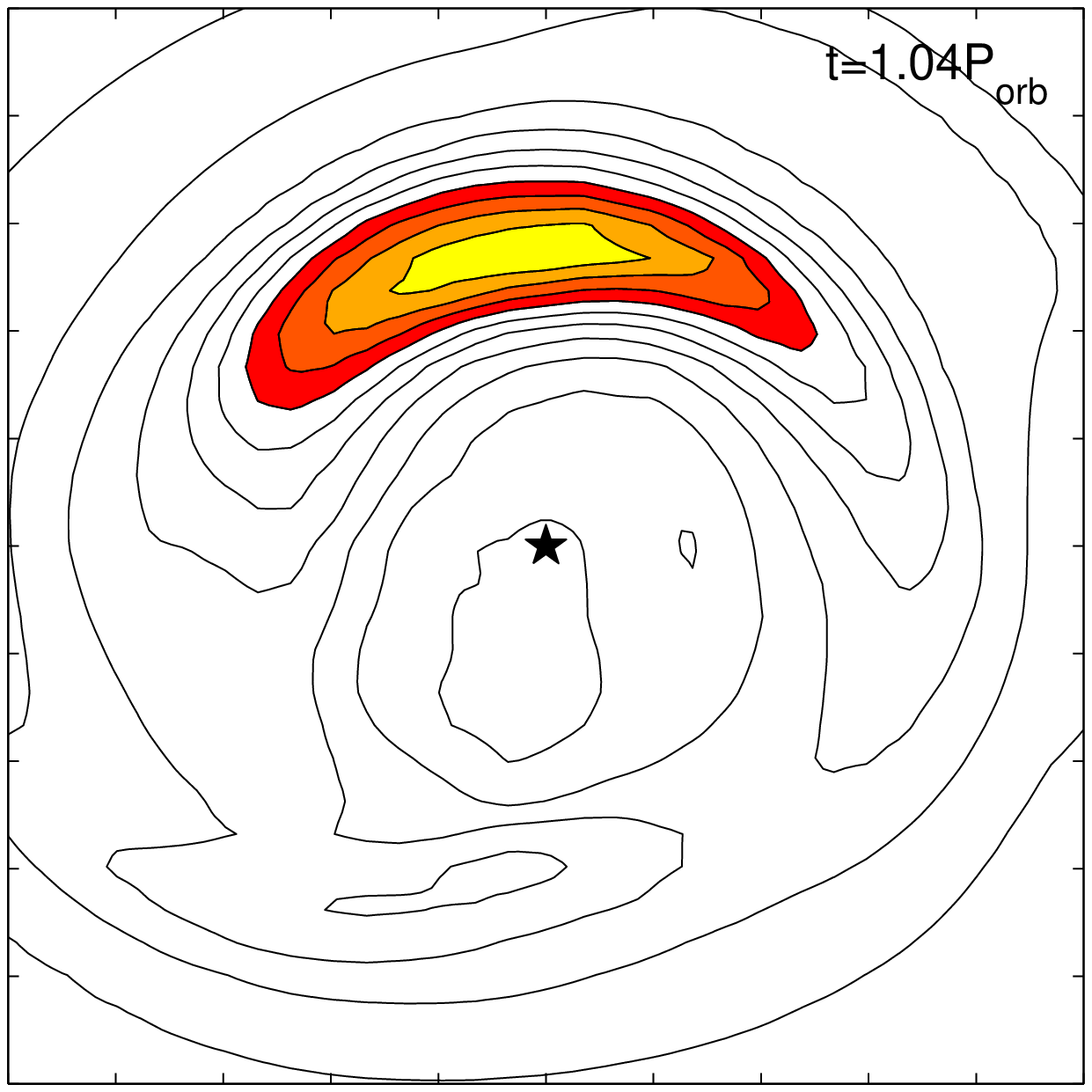,width=51.5mm}}}
\centerline{\hbox{\psfig{figure=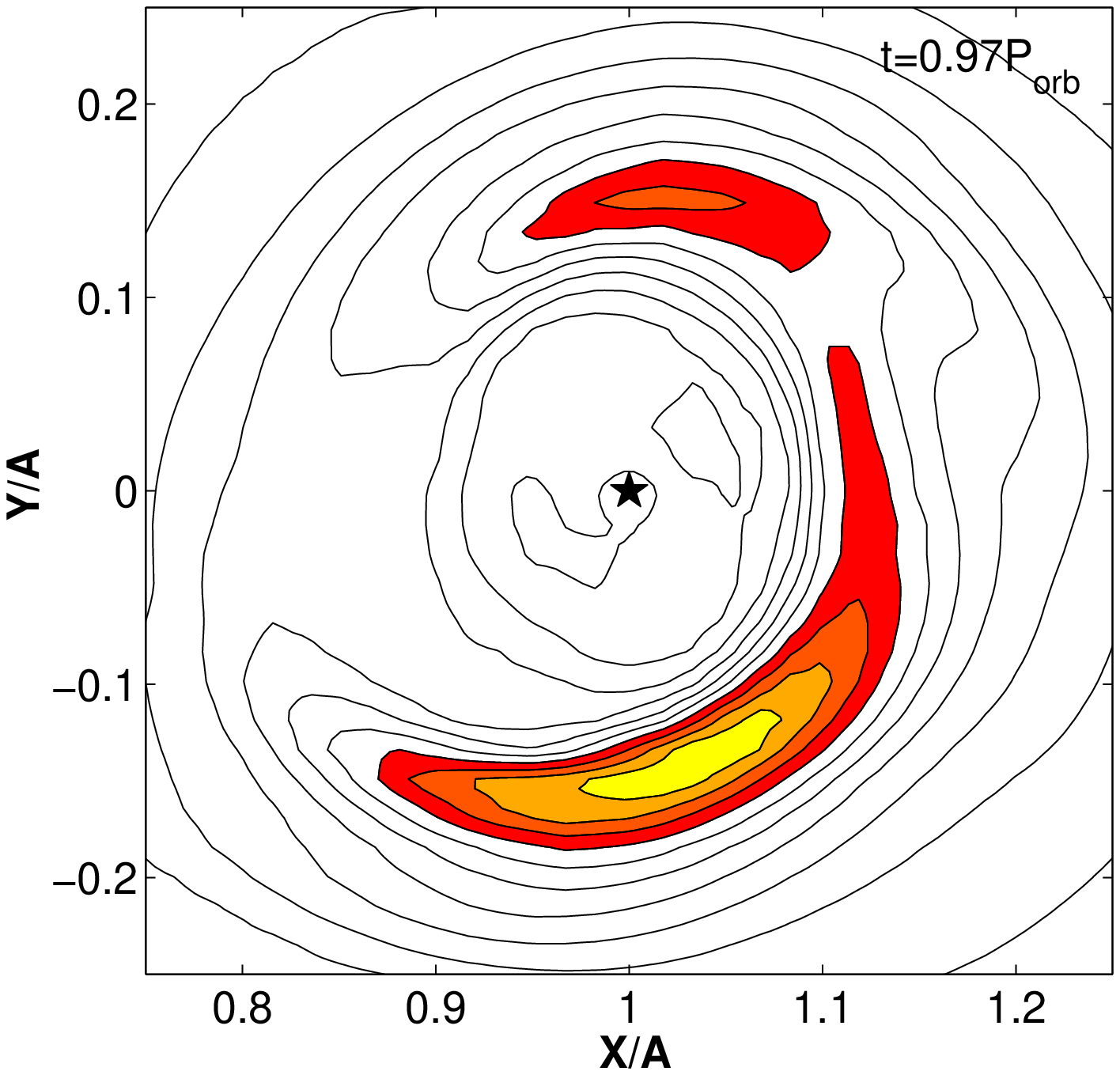,width=58mm}}
\hbox{\psfig{figure=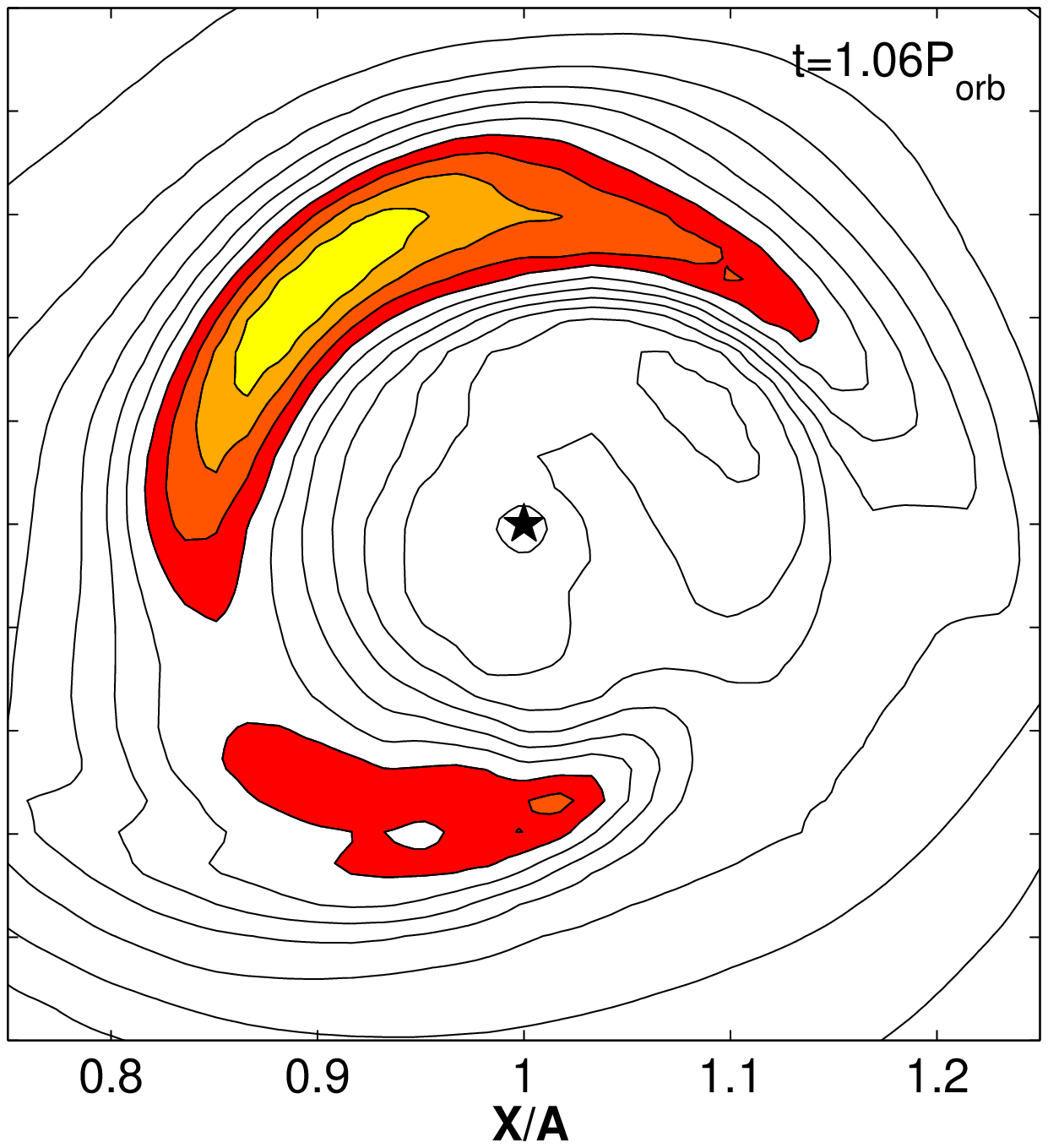,width=51.5mm}}}
\caption{\footnotesize Density distribution in the equatorial
plane for the second set of moments of time from Fig.~3.}
\end{figure*}

Our results also show the formation of dense blob in the residual
disk. The blob is formed after the stream vanishes, i.e. after
time \linebreak $\sim t_0+0.15P_{orb}$ and rotates with variable
velocity. The Figure 3 presents time variation of the mean
density of matter passing through a semi-plane $XZ$ ($Y=0$,
$X>A$) slicing the disk. Along with a general density fall we
can see here periodic oscillations called forth by passing the
blob. The blob doesn't smear out under the action of dissipation
and the period of its rotation ($\sim 0.18P_{orb}$) remains the
same practically up to the moment when the disk vanishes. Figure
3 also contains a zoom for the mean density variation for times
from $t_0$ to $t_0+1.4P_{orb}$. Here we marked 2 sets of moments
of time (8 moments per set), each set covering one period of the
mean density variation. The first set corresponds to the
moments of time $t_0+0.11P_{orb} \div t_0+0.26P_{orb}$, and the
second -- to $t_0+0.91P_{orb} \div t_0+1.06P_{orb}$. Figures~4a
and 4b show the density distributions for the moments of time of
these two sets. Analysis of the data of Fig.~4a,b, shows that
the blob retains by its interaction with the arms if spiral
shocks.

Let us consider this mechanism in detail. Initially appearing as
the residue of the stream, the blob tends to distribute
uniformly along the disk under the action of dissipation but
retards after passing trough spiral shock so the compactness
of the blob retains. Later the increasing density contrast
between the blob and the disk and growths of pressure gradient
will force the standstilled blob to move. When reaching the
second arm of spiral shock the process of formation/retainment
of the blob is repeated. This mechanism is confirmed by data of
Fig.~5.  Figure 5a presents the velocity of the blob (the
velocity of the point with maximal density) versus time (solid
line). The keplerian velocity of this point $V_{K}=\sqrt{GM_1/r}$
is also shown by dashed line. The velocity variations are shown
for time interval $t_0+0.91P_{orb} \div t_0+1.11P_{orb}$
including the time interval for Fig.~4b and covering the full
period of blob's revolution. Figure 5b shows variation of maximal
density in the equatorial plane for the same time interval. It is
seen from Fig.~5 that matter is retarded after passing the shock
and becomes more dense, the maximal density $\rho_{max}$ being
increased 1.5 times. When accumulating a sufficient amount of
matter, the blob comes off the shock and its velocity increases
while maximal density decreases. After passing the second arm of
spiral shock the blob is retarded again. Note that the contrast
of density between the blob and the disk remains the same up to
the moment when the blob disappears. In this moment of time the
disk is contracted and spiral shock disappears as well. Later
the density contrast blob/disk decreases under the action of
dissipation but the blob exists up to the full vanishing of the
disk. This is illustrated in Fig.~6 where time dependency of the
density contrast blob/disk is shown. The large value of density
contrast as well as the variable velocity of the blob revolution
(but with constant period $\sim0.18P_{orb}$) make this feature
very interesting for observations.

\renewcommand{\thefigure}{5}
\begin{figure}[p]
\centerline{\hbox{\psfig{figure=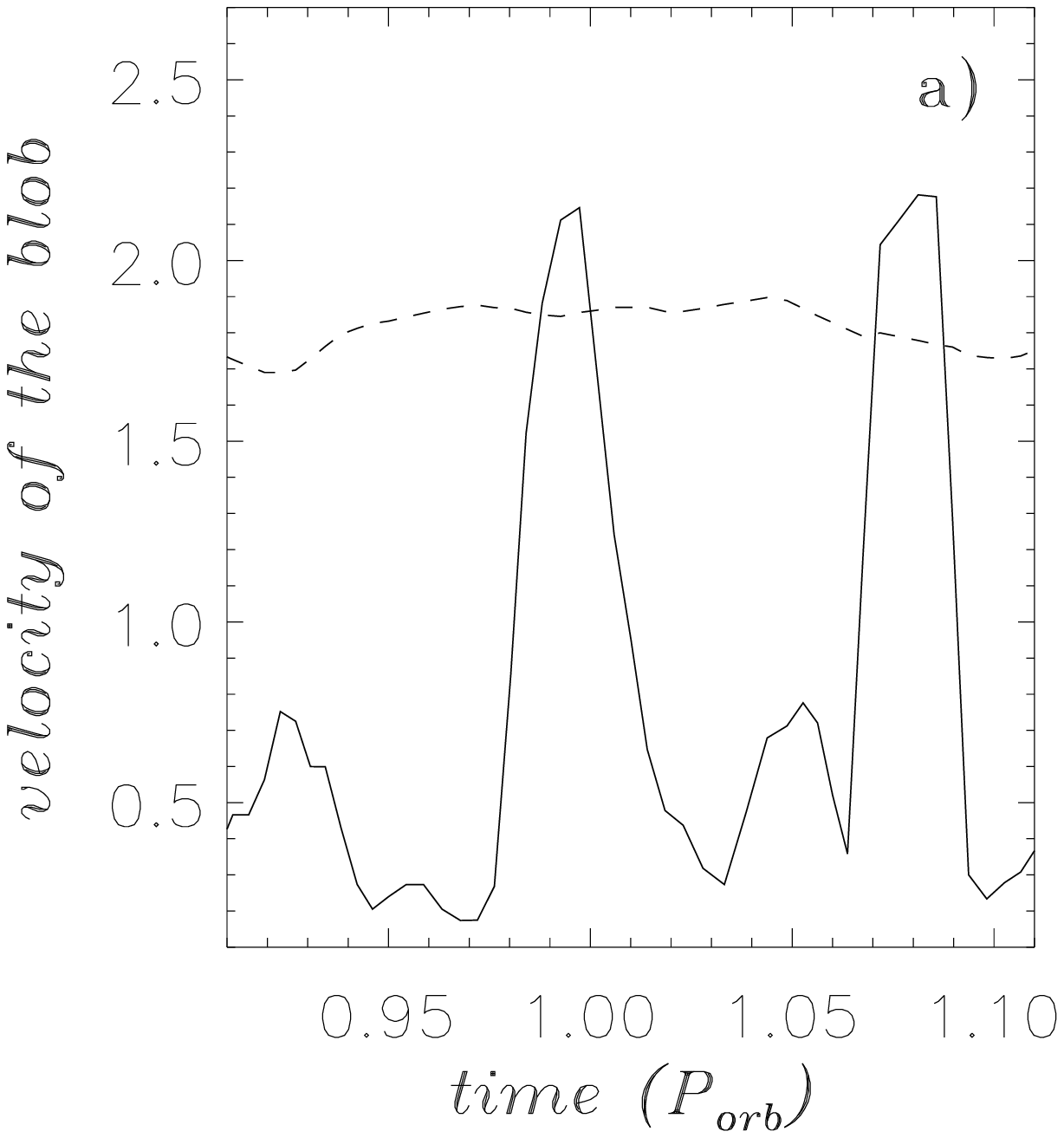,width=7cm}}}
\vspace*{1cm}
\centerline{\hbox{\psfig{figure=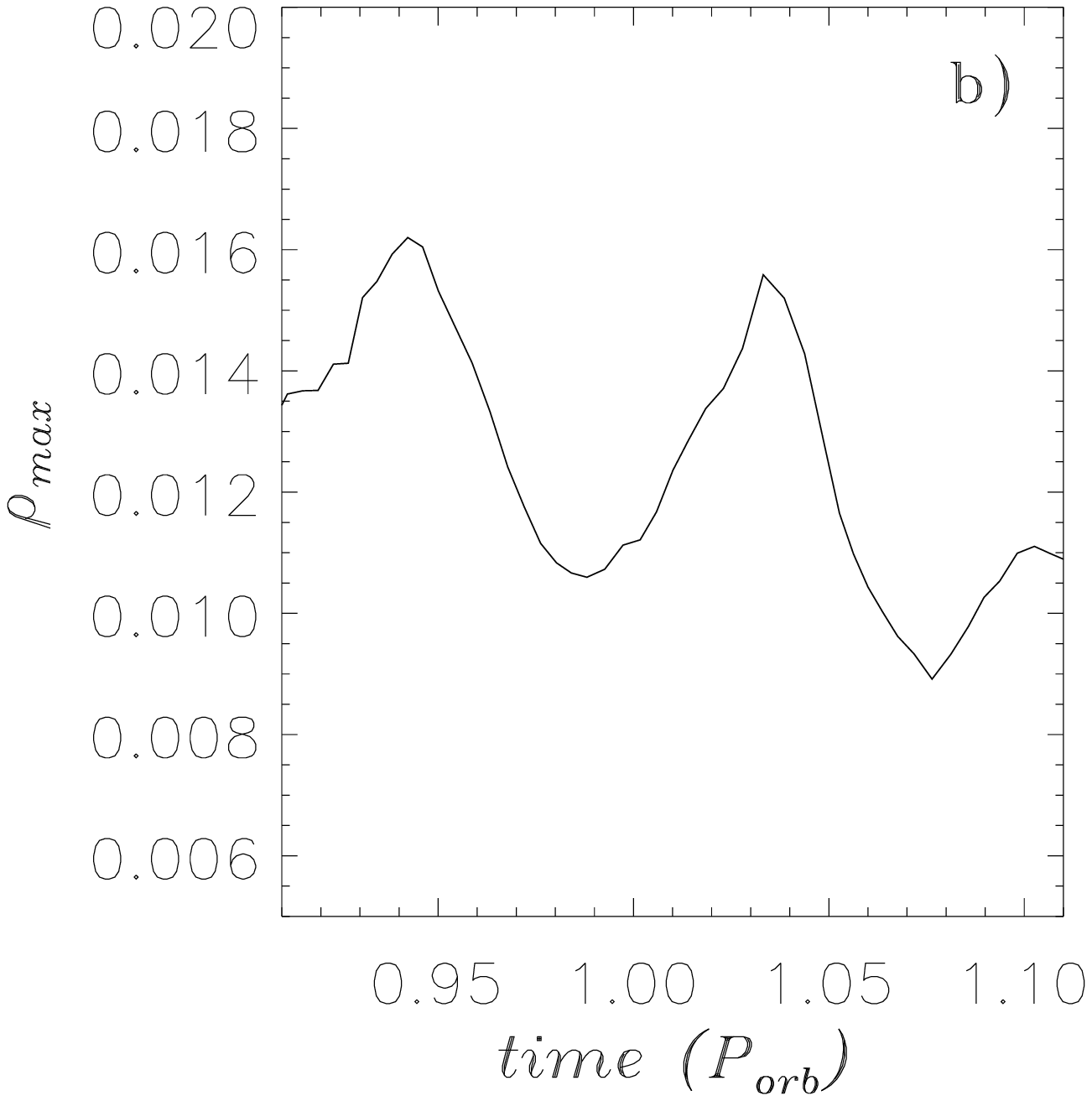,width=7cm}}}
\caption{\footnotesize {\normalsize\it Upper panel:} A solid
line -- variation of the velocity of the blob (the velocity
of the point with maximal density). A dashed line -- keplerian
velocity of this point $V_{K}=\protect\sqrt{GM_1/r}$. The
velocity variations are shown for time interval $t_0+0.91P_{orb}
\div t_0+1.11P_{orb}$ including the time interval for Fig.~4b
and covering the full period of blob's
revolution.\protect\\[3mm]
{\normalsize\it Lower panel:}
Variation of the maximal density of the blob.} \end{figure}

\renewcommand{\thefigure}{6}
\begin{figure}[t]
\centerline{\hbox{\psfig{figure=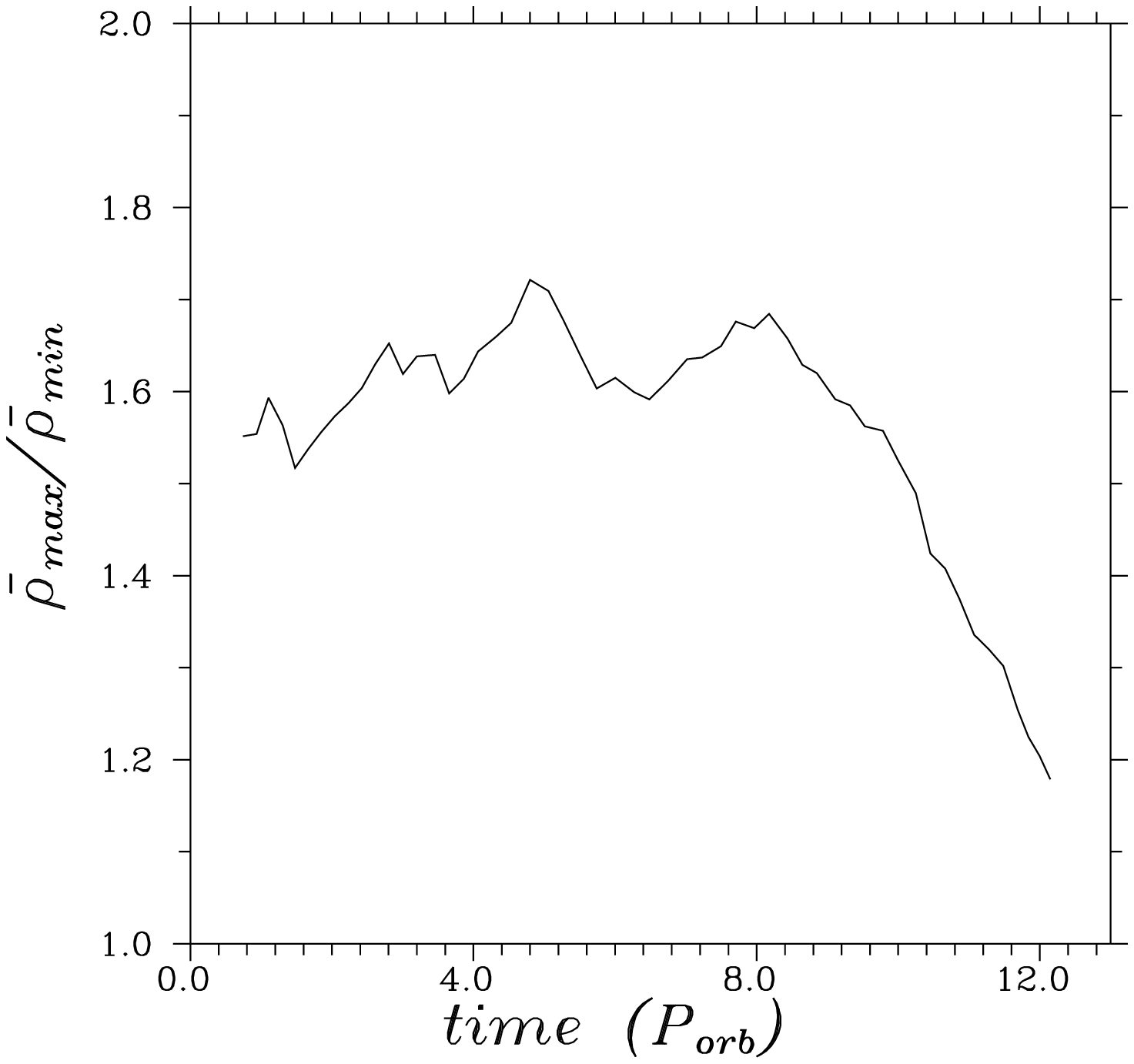,width=7cm}}}
\caption{\footnotesize The density contrast between the disk and
the blob versus time.}
\end{figure}

\section{Conclusions}

We have presented the results of 3D numerical simulations of
mass transfer in semidetached binaries after the mass transfer
termination.  Prior to simulation of flow structure with
`turned-off' mass transfer the near-steady-state solutions for
the case of constant non-zero rate of mass transfer was obtained
and used as the initial conditions. At the moment of time
$t=t_0$ the rate of mass transfer was decreased in five order of
magnitude, which corresponds to the cessation of mass transfer.
To investigate the influence of the viscosity we conduct 3 runs
for various values of viscosity, corresponding to following
values of parameter $\alpha$ (in terms of $\alpha$-disk):
$\alpha\sim0.08\div0.1$, $\alpha\sim0.04\div0.06$, and
$\alpha\sim0.01\div0.02$.

The investigation of structure of the residual disk reveals an
essential increasing of its lifetime when the viscosity
deceases. Even for $\alpha\sim0.05$ the lifetime of residual
disk exceeds $12P_{orb}$, and for $\alpha\sim0.01$ (that is
typical for observable accretion disks) the lifetime of residual
disk is as much as $50P_{orb}$.

Our simulations show that near the moment of time
$t=t_0+0.2P_{orb}$ the flow structure is changed significantly.
The stream from $L_1$ doesn't dominate anymore, and the shape
of accretion disk changes from quasi-elliptical to circular. The
second arm of tidally induced spiral shock is formed while
earlier (before the termination of mass transfer) it was
suppressed by the stream from $L_1$. The mass of the disk is
gradually decreased but spiral shocks exist practically up to
the full vanishing of the disk.

Our simulations also show that a dense blob is formed in the
residual disk, the velocity of the blob motion through the disk
being variable. The blob doesn't smear out under the action of
dissipation but is sustained by interaction with arms of spiral
shock up to the moment of time when spiral shock disappears. The
density contrast between the blob and the disk is rather large
$\sim1.6$ and begins to decrease only when spiral shock
disappears (it occurs at time $\sim t_0+10P_{orb}$) but even for
this time and up to the full vanishing of the residual disk the
density contrast is of order 1.2.

\section*{Acknowledgments}

The work was partially supported by Russian Foundation for Basic
Research and by grants of President of Russia.


\begin{thebibliography}{9999}

\raggedright

\footnotesize

\bibitem{armi96}
Armitage P.J., Livio M. 1996, Accretion Disks in Interacting
Binaries: Simulation of the Stream--Disk Impact, {\it Astrophys.
J.}, {\bf 470}, 1024

\bibitem{bath81}
Bath G.T., Pringle J.E. 1981, The Evolution of
Accretion Discs -- I. Mass Transfer Variations, {\it Monthly
Notices Roy. Astron. Soc.}, {\bf 194}, 967

\bibitem{bath83}
Bath G.T., van Paradijs J. 1983, Outburst Period--Energy
Relations in Cataclysmic Novae, {\it Nature}, {\bf 305}, 33

\bibitem{bath74}
Bath G.T., Evans W.D., Papaloizou J., Pringle J.E. 1974, The
Accretion Model of Dwarf Novae with Application to Z
Chameleontis, {\it Monthly Notices Roy. Astron. Soc.}, {\bf
169}, 447

\bibitem{dima97}
Bisikalo~D.V., Boyarchuk A.A., Kuznetsov~O.A., Chechetkin~V.M.
1997, Three-dimensional Modeling of the Matter Flow Structure in
Semidetached Binary Systems, {\it Astron. Zh.}, {\bf 74}, 880
({\it Astron. Reports}, {\bf 41}, 786, preprint
astro-ph/9802004)

\bibitem{dima98a}
Bisikalo~D.V., Boyarchuk A.A., Kuznetsov~O.A., Khruzina T.S.,
Cherepashchuk A.M.,  Chechetkin~V.M. 1998a, Evidence for the
Absence of a Stream-Disk Shock Interaction in Semi-Detached
Binary Systems: Comparison of Mathematical Modeling Results and
Observations, {\it Astron. Zh.}, {\bf 75}, 40 ({\it Astron.
Reports}, {\bf 42}, 33, preprint astro-ph/9802134)

\bibitem{dima98b}
Bisikalo~D.V., Boyarchuk A.A., Kuznetsov~O.A., Chechetkin~V.M.
1998b, The Influence of Parameters on the Flow Structure in
Semidetached Binary Systems: 3D Numerical Modeling, {\it Astron.
Zh.}, {\bf 75}, 706 ({\it Astron. Reports}, {\bf 42}, 621,
preprint astro-ph/9806013)

\bibitem{dima98c}
Bisikalo~D.V., Boyarchuk A.A., Chechetkin~V.M., Kuznetsov~O.A.,
Molteni~D. 1998c, 3D Numerical Simulation of Gaseous Flows
Structure in Semidetached Binaries, {\it Monthly Notices Roy.
Astron. Soc.}, {\bf 300}, 39

\bibitem{dima99}
Bisikalo~D.V., Boyarchuk A.A., Chechetkin~V.M., Kuznetsov~O.A.,
Molteni~D. 1999, Comparison of 2D and 3D Models of Flow
Structure in Semidetached Binaries, {\it Astron. Zh.}, {\bf 76},
905 ({\it Astron. Reports}, {\bf 43}, 797, preprint
astro-ph/9907084)

\bibitem{dima2000}
Bisikalo~D.V., Boyarchuk A.A., Kuznetsov~O.A., Chechetkin~V.M.
2000, The Impact of Viscosity on the Flow Structure Morphology
in Semidetached Binary Systems. Results of 3D Numerical
Modeling, {\it Astron. Zh.}, {\bf 77}, 31 ({\it Astron.
Reports}, {\bf 44}, 26, preprint astro-ph/9907087)

\bibitem{osher85}
Chakravarthy S.R., Osher S. 1985, A New Class of High Accuracy
TVD Schemes for Hyperbolic Conservation Laws, {\it AIAA Pap.}, N
85-0363

\bibitem{einfeldt88}
Einfeldt B. 1988, On Godunov-Type Methods for Gas Dynamics, {\it
SIAM J. Numer. Anal.}, {\bf 25}, 294

\bibitem{gilliland85}
Gilliland R.L. 1985, Hydrodynamical Modeling of Mass Transfer
from Cataclysmic Variable Secondaries, {\it Astrophys. J.},
{\bf 292}, 522

\bibitem{lubowshu75}
Lubow S.H., Shu F.H. 1975, Gas Dynamics of Semidetached
Binaries, {\it Astrophys. J.}, {\bf 198}, 383

\bibitem{lyndenbell74}
Lynden-Bell~D., Pringle J.E. 1974, The Evolution of Viscous
Discs and the Origin of Nebular Variables, {\it Monthly Notices
Roy. Astron. Soc.}, {\bf 168}, 603

\bibitem{meyer93}
Meyer-Hofmeister E., Ritter H. 1993, Accretion Disks in Close
Binaries, in "The Realm of Interacting Binary Stars", eds
J.Sahade, G.E.McCluskey,Jr., Y.Condo, Dordrecht: Kluwer Academic
Publishers, p.143

\bibitem{diego91}
Molteni~D., Belvedere G., Lanzafame G. 1991, Three-dimensional
Simulation of Polytropic Accretion Discs, {\it Monthly Notices
Roy. Astron. Soc.}, {\bf 249}, 748

\bibitem{murray2000}
Murray J.R., Warner B., Wickramasinghe~D.T. 2000, Eccentric
Discs in Binary with Intermediate Mass Ratios:  Superhumps in
the VY Sculpturis Stars, {\it Monthly Notices Roy. Astron.
Soc.}, {\bf 315}, 707

\bibitem{tanya2001}
Khruzina T.S., Cherepashchuk A.M., Bisikalo~D.V., Boyarchuk
A.A., Kuznetsov~O.A. 2001, The Interpretation of Light Curves of
IP Peg in the Model of Shock-free Interaction between the Gas
Stream and the Disk, {\it Astron. Zh.}, in press

\bibitem{ritter88}
Ritter H. 1988, Turning On and Off Mass Transfer in Cataclysmic
Binaries, {\it Astron. Astrophys.}, {\bf 202}, 93

\bibitem{roe86}
Roe P.L. 1986, Characteristic-Based Schemes for the Euler
Equations, {\it Ann. Rev. Fluid Mech.}, {\bf 18}, 337

\bibitem{spiral1}
Sawada K., Matsuda T., Hachisu I. 1986, Spiral Shocks on a Roche
Lobe Overflow in Semidetached Binary System, {\it Monthly
Notices Roy. Astron. Soc.}, {\bf 219}, 75

\bibitem{spiral2}
Sawada K., Matsuda T., Hachisu I. 1986, Accretion Shocks in a
Close Binary System, {\it Monthly Notices Roy. Astron. Soc.},
{\bf 221}, 679

\bibitem{sawada87}
Sawada K., Matsuda T., Inoue M., Hachisu I. 1987, Is the
Standard Accretion Disc Model Invulnerable? {\it Monthly Notices
Roy. Astron. Soc.}, {\bf 224}, 307

\bibitem{schreiber2000}
Schreiber~M.R., G\"ansicke~B.T., Hessman~F.V. 2000, The Response
of a Dwarf Nova Disc to Real Mass Transfer Variations, {\it
Astron. Astrophys.}, {\bf 358}, 221

\bibitem{tout96}
Tout C. 1996, Accretion Disk Viscosity, in "Cataclysmic
Variables and Related Objects", eds A.Evans, J.H.Wood,
Dordrecht: Kluwer Academic Publishers, p.97

\bibitem{wood77}
Wood P.R. 1977, Mass Transfer Instabilities in Binary Systems,
{\it Astrophys. J.}, {\bf 217}, 530

\end{thebibliography}
\end{document}